\documentstyle[12pt,equation]{article}
\begin{document}

\newcommand{\sigmin}{\mbox{$\sigma_{\gamma \gamma}^{\rm min}$}}
\newcommand{\as}{\mbox{$\alpha_s$}}
\newcommand{\amu}{\mbox{$A_t + \mu \cot \! \beta$}}
\newcommand{\mgg}{\mbox{$M_{\gamma \gamma}$}}
\newcommand{\fourl}{\mbox{$l^+ l^+ l^- l^-$}}
\newcommand{\fourtau}{\mbox{$\tau^+ \tau^+ \tau^- \tau^-$}}
\newcommand{\tht}{\mbox{$\theta_t$}}
\newcommand{\tanb}{\mbox{$\tan \! \beta$}}
\newcommand{\msb}{\mbox{$m_{\tilde{b}_L}$}}
\newcommand{\msbsq}{\mbox{$m^2_{\tilde{b}_L}$}}
\newcommand{\mt}{\mbox{$m_t$}}
\newcommand{\mtsq}{\mbox{$m^2_t$}}
\newcommand{\mst}{\mbox{$m_{\tilde{t}_1}$}}
\newcommand{\mstt}{\mbox{$m_{\tilde{t}_2}$}}
\newcommand{\mstl}{\mbox{$m_{\tilde{t}_L}$}}
\newcommand{\mstr}{\mbox{$m_{\tilde{t}_R}$}}
\newcommand{\mstlsq}{\mbox{$m^2_{\tilde{t}_L}$}}
\newcommand{\mstrsq}{\mbox{$m^2_{\tilde{t}_R}$}}
\newcommand{\mstlr}{\mbox{$m_{\tilde{t}_{L,R}}$}}
\newcommand{\mstlrsq}{\mbox{$m^2_{\tilde{t}_{L,R}}$}}
\newcommand{\mstsq}{\mbox{$m^2_{\tilde{t}_1}$}}
\newcommand{\msttsq}{\mbox{$m^2_{\tilde{t}_2}$}}
\newcommand{\stst}{\mbox{$\tilde{t}_1 \tilde{t}_1^*$}}
\newcommand{\sigst}{\mbox{$\sigma_{\tilde{t}_1}$}}
\newcommand{\msig}{\mbox{$m_{\sigma_{\tilde t}}$}}
\newcommand{\gamgam}{\mbox{$\gamma \gamma$}}
\newcommand{\st}{\mbox{$\tilde{t}_1$}}
\newcommand{\stt}{\mbox{$\tilde{t}_2$}}
\newcommand{\stl}{\mbox{$\tilde{t}_L$}}
\newcommand{\str}{\mbox{$\tilde{t}_R$}}
\newcommand{\ww}{\mbox{$W^+W^-$}}
\newcommand{\epem}{\mbox{$e^+e^-$}}
\newcommand{\mat}{\mbox{${\cal M}^2_{\tilde{t}}$}}
\newcommand{\be}{\begin{equation}}
\newcommand{\ee}{\end{equation}}
\newcommand{\een}{\end{subequations}}
\newcommand{\ben}{\begin{subequations}}
\newcommand{\beq}{\begin{eqalignno}}
\newcommand{\eeq}{\end{eqalignno}}
\renewcommand{\thefootnote}{\fnsymbol{footnote} }
\noindent
\begin{flushright}
MAD/PH/808\\
KEK--TH--379\\
KEK Preprint 93--163\\
November 1993
\end{flushright}
\vspace{1.5cm}
\pagestyle{empty}
\begin{center}
{\Large \bf Production and Decay of Scalar Stoponium Bound States}\\
\vspace*{5mm}
Manuel Drees\footnote{Heisenberg Fellow}\\
{\em Physics Department, University of Wisconsin, Madison, WI 53706, USA}\\
Mihoko M. Nojiri\footnote{E--mail: NOJIRIN@JPNKEKVX}\\
{\em Theory Group, KEK, Oho 1--1, Tsukuba, Ibaraki 305, Japan}
\end{center}

\begin{abstract}
In this paper we discuss possible signatures for the production of scalar
\stst\ (stoponium) bound states \sigst\ at hadron colliders, where \st\ is the
lighter scalar top eigenstate. We first study the decay of \sigst; explicit
expressions are given for all potentially important decay modes. If \st\ has
unsuppressed two--body decays, they will always overwhelm the annihilation
decays of \sigst. Among the latter, we find that usually either the $gg$ or
$hh$ final state dominates, depending on the size of the off--diagonal entry
of the stop mass matrix; $h$ is the lighter neutral scalar Higgs boson of the
minimal supersymmetric model. If \msig\ happens to be close to the mass of one
of the neutral scalar Higgs bosons, $Q \bar{Q}$ final states dominate ($Q=b$
or $t$). \ww\ and $ZZ$ final states are subdominant. We argue that $\sigst
\rightarrow \gamgam$ decays offer the best signal for stoponium production at
hadron colliders. The tevatron should be able to close the light stop window
left open by LEP searches, but its mass reach is limited to $\msig \leq 90$
GeV. In contrast, at the LHC one should ultimately be able to probe the region
$\msig \leq 700$ GeV, if the $hh$ partial width is not too large. We also
comment on the feasibility of searching for \sigst\ production at hadron
colliders in the $ZZ, \ Z \gamma$ and \fourtau\ final states, and briefly
mention \sigst\ production at \gamgam\ colliders.
\end{abstract}
\clearpage
\setcounter{footnote}{0}
\pagestyle{plain}
\setcounter{page}{1}
\section*{1) Introduction}
Although the Standard Model (SM) of Particle Physics \cite{1} has so far
passed all experimental tests, it has long been known \cite{2} to be
technically unnatural: Nothing protects the mass of the Higgs boson, and hence
the scale of electroweak symmetry breaking, from large (quadratically
divergent) radiative corrections which ``naturally" push it up to the Planck
scale or the scale $M_X$ of Grand Unified Theories (GUTs). The probably most
elegant solution \cite{3} of this problem is the introduction of $N=1$
supersymmetry (SUSY) \cite{4}. In supersymmetric theories corrections to the
mass of the Higgs boson from loops involving SM particles are automatically
cancelled by loops involving their superpartners. The cancellation is not
perfect since supersymmetry has to be broken; naturalness arguments then
suggest that the scale of SUSY breaking should not (much) exceed 1 TeV.

So far searches for the direct production of superparticles (sparticles)
in collider experiments have not been successful. However, the lower limits
on their masses that can be inferred from these searches are only around 120
GeV for most strongly interacting sparticles \cite{5}, and under 50 GeV for
all other sparticles \cite{6}. This leaves a wide region to be explored by
present and future experiments, and much work in that direction has already
been done \cite{4,7}.

Meanwhile various indirect (loop) effects due to supersymmetric particles
have been investigated. A by now well--known result \cite{8} is that the
introduction of supersymmetry allows for a beautiful unification of all three
gauge couplings of the SM at scale $M_X \simeq 10^{16}$ GeV. In contrast,
nonsupersymmetric theories can be unified only at the cost of the somewhat
ad hoc introduction of new degrees of freedom and/or intermediate scales
between $M_X$ and $M_W$. Unfortunately these analyses only tell us \cite{8,9}
that the scale of SUSY breaking should not exceed 10 TeV or so, and hence
offer no immediate clues where to look for more direct evidence for the
existence of supersymmetric particles.

Such a clue might come from the third main motivation for the introduction of
SUSY. In addition to technical naturalness and simple Grand Unification,
supersymmetric theories also offer the possibility to {\em understand} (as
opposed to parametrize) electroweak gauge symmetry breaking, in terms of
(logarithmic) radiative corrections to scalar masses. Even if all these
masses are identical at some very high energy scale where SUSY breaking
becomes effective, as e.g. in minimal Supergravity theories \cite{4},
radiative corrections will drive the square of the mass of one Higgs boson
doublet to negative values at low energies, leaving all other squared scalar
masses positive \cite{10}. The driving force in this radiative symmetry
breaking is the large Yukawa coupling of the top quark: Radiative
corrections due to Yukawa couplings reduce scalar masses, while gauge
interactions increase them. This mechanism not only establishes a causal
link between the breaking of supersymmetry and the breakdown of electroweak
gauge symmetry, it also points towards a fundamental role for a large (top)
Yukawa coupling, and might thus eventually help to understand why the top
quark is so much heavier than all other SM fermions.

In more practical terms, these considerations indicate that the superpartners
of the top quark, the stop squarks, might be considerably {\em lighter} than
the other squarks \cite{11}; recall that radiative corrections due to Yukawa
interactions {\em reduce} scalar masses. In addition a large Yukawa coupling
implies large mass mixing between the superpartners of left-- and
right--handed top quarks, which further reduces the mass of the lighter stop
eigenstate \st. As a result, even in minimal supergravity models \mst\ can be
almost arbitrarily light even if all other squarks have masses of several
hundred GeV \cite{12,13}.

What is the experimental situation? Stop squarks are color triplets, and thus
have substantial pair production cross sections at hadron colliders. However,
present experimental bounds \cite{5} on squark masses assume 10 or 12
degenerate squark eigenstates, and therefore do not apply to \st; at present,
searches for events with large missing transverse energy cannot exclude the
existence of a light stop if the mass of the lightest supersymmetric particle
(LSP) exceeds 12 GeV or so \cite{14}. Stop squarks also have electromagnetic
and weak interactions. However, for a certain \stl--\str\ mixing angle \tht\
the $Z \stst$ coupling vanishes \cite{15}, so that {\em no} strict lower bound
on \mst\ can be derived from the study of $Z$ decays. In this scenario \st\
pair production at \epem\ colliders can only proceed via photon exchange. The
lower bound on \mst\ will then again depend on the LSP mass, since for too
small a \st\ -- LSP mass difference \st\ pair events look more and more like
hadronic two--photon events.\footnote{The TOPAZ collaboration at the TRISTAN
collider recently observed \cite{topaz} a slight excess of $D^*$ mesons when
trying to measure two--photon production of charm. This has caused
speculations \cite{eno,13,21} that there might be a stop squark with mass
between 15 and 20 GeV; \stst\ production would then give a similar signal as
two--photon production of charm if the LSP lies just a few GeV below the stop.
However, the QCD prediction for $\sigma(\gamgam \rightarrow c \overline{c})$
is quite uncertain \cite{16}; the measured cross section is only about 1.5
standard deviations above the upper range of QCD predictions. Moreover,
preliminary data from the VENUS collaboration \cite{venus} indicate that the
\st--LSP mass difference would have to be as small as 2.5 GeV for \st\ to have
escaped detection. Interpretation of the excess in terms of a rather contrived
SUSY model therefore seems premature.}

One might worry that a light stop squark produces unacceptably large loop
effects. Indeed, a light stop can lead to large corrections to the
electroweak $\rho$ parameter and related quantities \cite{17}. However, the
main effect would be a reduction of the value of the top quark mass $m_t$
fitted from electroweak data; note that at present the central value of the SM
fit of $m_t$ \cite{18} is substantially above the direct search limit on
$m_t$ \cite{19}. Moreover, this search limit is only valid in the SM. If \st\
is light enough, the decay $t \rightarrow \st \ +$ LSP can significantly
dilute the di--lepton signal for top, especially if $m_t \leq 90$ GeV
\cite{14}. In any case, a small \mst\ need not imply large contributions to
$\delta \rho$ \cite{17}. Similarly, loops involving stop squarks and charginos
can contribute significantly to the matrix element for $b \rightarrow s
\gamma$ decays \cite{20}. However, this contribution can be cancelled by other
non--SM loop contributions involving charged Higgs bosons. As a result, the
recent bounds \cite{cleo} on the branching ratio for $b \rightarrow s \gamma$
do not exclude \cite{21} a very light \st.

We thus conclude that a \st\ of 40 or 50 GeV could quite easily have escaped
detection so far; if we are willing to finetune the $Z \stst$ coupling
and the \st\ -- LSP mass difference even a 15 or 20 GeV \st\ is not excluded.
This also indicates that it is difficult to obtain stringent bounds on \mst\
from searches for open \st\ production. In particular, most signals will
disappear in the limit where \st\ becomes (almost) degenerate with the LSP;
unlike for, say, first generation squarks and gluinos this is possible even
in the restrictive class of minimal supergravity models
\cite{12,13}.\footnote{At least in models with exact $R-$parity \st\ cannot
actually be the LSP, since searches for exotic isotopes imply \cite{22} that
the LSP has to be electrically and color neutral.}

On the other hand, if the \st\ -- LSP mass difference is small, \st\ will be
rather long--lived \cite{23}. The reason is that in this case two--body
tree--level decays like $\st \rightarrow t \ +$ LSP or $\st \rightarrow b \ +
$ chargino are kinematically forbidden. In this situation \st\ decays
preferably into a charm quark and a neutralino via a loop diagram, whose
decay width is suppressed by a factor $\sim 10^{-7}$ compared to tree--level
decays \cite{23}. Being long--lived, stops can form bound states
(``stoponia"), which eventually decay via \stst\ annihilation into final
states that only contain SM particles.

In this paper we study the decay of scalar (S--wave) \stst\ bound states
\sigst\, as well as possible signals for \sigst\ production at hadron
(super)colliders. Scalar stoponium has been studied previously in
refs.\cite{24,25,26}. However, in refs.\cite{24} and \cite{26} mixing between
the superpartners of left-- and right--handed top quarks was ignored, and
diagrams involving Higgs bosons in the intermediate or final state were
treated only in an approximate fashion or not at all; both effects can be
very important. They have been included in ref.\cite{25}, but there only a
very light \sigst\ was treated, so that many decay channels were kinematically
forbidden. We computed the decay widths for all tree--level two--body decays
of \sigst\ for general stop mixing and the whole range of masses of interest
in the foreseeable future. As already pointed out in our recent Letter
\cite{27} the two--photon decay of \sigst\ probably offers the best signal at
hadron colliders. Here we present a more detailed discussion of the region of
parameter space where this signal is viable. We also point out that an
interesting range of \sigst\ masses should already be accessible at the
Fermilab Tevatron.

The rest of this article is organized as follows. In sec. 2 we discuss \sigst\
decays. Final states consisting of two gluons or two scalar Higgs bosons
usually are dominant among decays that proceed by \stst\ annihilation.
Contrary to claims in refs.\cite{24,26} final states containing two heavy
gauge bosons contribute much less to the total width of \sigst; this is a
direct consequence of electroweak gauge invariance. In sec. 3 we discuss
signals for \sigst\ production at present and future hadron colliders. We
focus on the clean two--photon signal, whose background can be calculated
relatively reliably. At future supercolliders a signal consisting of 4 $\tau$
leptons may also be viable, but here realistic background estimates are much
more difficult. Sec. 4 contains a summary of our main results and some
conclusions. Explicit expressions for all \sigst\ two--body decays are
collected in the Appendix.
\setcounter{footnote}{0}
\section*{2) Stoponium decays}
In this section we discuss the decays of scalar \stst\ bound states \sigst. We
work within the framework of the minimal supersymmetric standard model (MSSM)
\cite{4}, which is a supersymmetrization of the SM with minimal Higgs sector.
In particular, every quark gets two superpartners described by complex scalar
fields, which are associated with the left-- and right--handed components of
the quark Dirac spinor. If the quark is massive, its two superpartners can mix
with each other. For the case of stop squarks the resulting mass matrix is
given by \cite{11} (in the basis $(\stl,\str)$):
\be \label{e1}
\mat = \mbox{$ \left( \begin{array}{cc}
m^2_{\tilde{t}_L} + m_t^2 + 0.35 m_Z^2 \cos \! 2 \beta
& - m_t (A_t + \mu \cot \! \beta) \\
- m_t (A_t + \mu \cot \! \beta ) &
m^2_{\tilde{t}_R} + m_t^2 + 0.16 m_Z^2 \cos \! 2 \beta
\end{array} \right) $}, \ee
where we have used the conventions of ref.\cite{12}, which are quite similar
to those of ref.\cite{28}. The quantities \mstl, \mstr\ describe the soft SUSY
breaking contributions to the diagonal squark masses. As already mentioned in
the Introduction, loop corrections involving the top Yukawa coupling tend to
reduce these mass parameters compared to their values at higher energies. In
models where all squark masses are equal at some very high (GUT or Planck)
scale one therefore expects \mstlr\ to be {\em smaller} than the corresponding
quantities for squarks of the first two generations. The same argument also
implies $\mstr < \mstl$, since Yukawa interactions affect the running of
\mstr\ more strongly \cite{10}.

The diagonal entries of the stop mass matrix (\ref{e1}) also depend on the
mass \mt\ of the top quark as well as the angle $\beta$, defined via $\tanb
\equiv \langle H_2^0 \rangle / \langle H_1^0 \rangle $, where $H_1, \ H_2$ are
the two Higgs doublet fields necessary in any realistic supersymmetric model
\cite{4}. Two additional parameters enter the off--diagonal entries of
(\ref{e1}): The supersymmetric Higgs(ino) mass $\mu$, and the parameter $A_t$
describing the strength of the trilinear scalar $H_2^0 \stl \tilde{t}^*_R$
interaction, which breaks supersymmetry. $A_t$ and $\mu$ are free parameters
of the model, but we generally expect them to be of roughly the same magnitude
as \mstlr; \tanb\ can be taken to be positive, but both $A_t$ and $\mu$ can
have either sign.

The mass matrix (\ref{e1}) can easily be diagonalized. We define
\be \label{e2}
\mbox{$\left( \begin{array}{c}
\tilde{t}_1 \\ \tilde{t}_2 \end{array} \right) $} =
\mbox{$\left( \begin{array}{cc}
\cos \! \theta_t & \sin \! \theta_t \\
-\sin \! \theta_t & \cos \! \theta_t \end{array} \right) $}
 \mbox{$\left( \begin{array}{c}
\tilde{t}_L \\ \tilde{t}_R \end{array} \right) $}, \ee
and obtain \ben \label{e3} \beq
m^2_{\tilde{t}_{1,2}} &= \frac{1}{2} \left[ m^2_{LL} + m^2_{RR} \mp
\sqrt{ \left( m^2_{LL} - m^2_{RR} \right)^2 + 4 m^4_{LR} } \right];
\label{e3a} \\
\tan \! \theta_t &= \frac {m^2_{\tilde{t}_1} - m^2_{LL} } {m^2_{LR}},
\label{e3b} \eeq \een
where $m^2_{LL,LR,RR}$ are the (1,2), (1,2) and (2,2) elements of the mass
matrix (\ref{e1}), respectively. Note that $\cos^2 \theta_t < \ (>) \ 1/2$ if
$\mstlsq > \ (<) \ \mstrsq$.

We will see below that \sigst\ decays can involve charginos and neutralinos as
well as Higgs bosons; we therefore briefly describe the corresponding sectors
of the MSSM. The charginos $\tilde{W}_1, \ \tilde{W}_2$ are \cite{4} mixtures
of the superpartners of the $W^{\pm}$ bosons and of the charged Higgs fields
contained in $H_{1,2}$. Similarly, the neutralinos $\tilde{Z}_i, \
i=1,\dots,4$ are mixtures of the superpartners of the (unmixed) $B$ and $W_3$
gauge bosons of the SM as well as of the neutral Higgs bosons. Charginos are
Dirac fields while neutralinos can be described by Majorana spinors. The mass
matrices for charginos and neutralinos depend \cite{4} on the parameters $\mu$
and \tanb\ introduced above, as well as on the SUSY breaking gaugino masses
$M_1$ and $M_2$. We will for simplicity assume the usual unification relation
$M_1 = 5/3 \tan^2 \theta_W M_2 \simeq M_2/2$; similarly, the gluino mass is
given by $M_3 = \alpha_s \sin^2 \theta_W M_2 / \alpha$. The description of the
neutralino, chargino and gluino sectors thus only necessitates the
introduction of one additional parameter, which we take to be the mass $M_2$
(at the weak scale). Finally, at the tree level the Higgs sector of the MSSM
is described by two parameters \cite{28}, which we take to be \tanb\ and the
mass $m_P$ of the pseudoscalar Higgs boson. We have included radiative
corrections to the Higgs masses and mixings from top--stop loops \cite{29},
employing the effective potential formalism \cite{30}; these corrections are
determined completely in terms of the parameters entering the stop mass matrix
(\ref{e1}).

We are now in a position to discuss \sigst\ decays. There are two very
different kinds of decay modes: {\em Single stop decays}, where one stop squark
decays independently of the second squark; and {\em  annihilation decays},
which proceed by \stst\ annihilation. The \sigst\ partial widths
due to single stop decays are simply twice the corresponding stop decay
widths. Since we are interested in relatively light stop squarks, we will
assume that $\st \rightarrow t + \tilde{g}$ decays are kinematically
forbidden. We then have to consider the following two--body \st\ decays:
\ben \label{e4} \beq
\st &\rightarrow b \tilde{W}_i, i=1,2 ; \label{e4a} \\
\st &\rightarrow t \tilde{Z}_j, j=1,\dots,4 ; \label {e4b} \\
\st &\rightarrow c \tilde{Z}_j, j=1,\dots,4 . \label {e4c}
\eeq \een
The processes (\ref{e4a},b) occur at tree level, with full gauge or
top Yukawa strength. If they are kinematically unsuppressed, the corresponding
\st\ decay widths are very roughly of order $10^{-3}$ to $10^{-2} \mst$;
exact expressions are listed in the Appendix, eqs.(\ref{ap15},\ref{ap16}).

If the channels (\ref{e4a},b) are kinematically closed, (\ref{e4c})
is usually the dominant decay mode \cite{23} of \st. Note that it couples a
third generation squark to a second generation quark, via a neutral
(super)current. In models where the squark mass matrix and the quark Yukawa
coupling matrix commute at some energy scale, e.g. if all squarks are mass
degenerate at some energy scale, the process (\ref{e4c}) therefore has to
proceed via a weak loop diagram involving small elements of the quark flavor
mixing matrix. Even though the amplitude is enhanced by a factor
$\log(M_X/M_W)$ the authors of ref.\cite{23} therefore estimate the squared
matrix element for the process (\ref{e4c}) to be suppressed by a factor $\sim
10^{-7}$ compared to the tree--level process (\ref{e4b}), leading to a decay
width $\sim 10^{-9} \mst$. We will see below that this is completely
negligible compared to the widths for annihilation decays, to which we turn
next.

As already mentioned, these proceed via the annihilation of the \st\ and
$\tilde{t}_1^*$ that make up \sigst; this kind of decay by far dominates the
total widths of the  familiar lowest lying quarkonium states $(\eta_c, J/\psi,
\Upsilon)$. Here we are only interested in two--body decays of \sigst, which
dominate all other annihilation decays. We treat \sigst\ as a nonrelativistic
bound state, where the squarks are in an $S-$wave. The partial width for
$\sigst \rightarrow A + B$ is then given by \cite{31,32}: \be \label{e5}
\Gamma \left( \sigst \rightarrow A + B \right) = \frac { 3 \beta}
{32 \pi^2 m^2_{\sigma_{\tilde t}} } | R(0) |^2 \frac {1} {1 + \delta_{AB} }
\sum_{\lambda_A, \lambda_B} | {\cal M}^{\lambda_A, \lambda_B} (\stst
\rightarrow A B) |^2_{v=0} . \ee
Here, \be \label{e6}
\beta = \sqrt{ \left( 1 - \frac {m_A^2 + m_B^2} {m^2_{\sigma_{\tilde t}} }
\right)^2
- \frac {4 m_A^2 m_B^2} {m^4_{\sigma_{\tilde t }} } } \ee
is the usual phase space factor, and $1/(1 + \delta_{AB})$ is a statistics
factor. Finally, $R(0)$ is the wave function at the origin. For realistic
QCD potentials the wave function generally has to be computed numerically.
Ref.\cite{33} provides parametrizations for $R(0)$ as well as the binding
energy of the first ten $S-$wave states of a nonrelativistic (s)quarkonium
system, using a potential that describes the known $c \overline{c}$ and $b
\overline{b}$ systems well. We will use their parametrizations throughout.

Eq.(\ref{e5}) reduces the problem of computing \sigst\ annihilation decay
widths to the calculation of the Feynman amplitude ${\cal M}^{\lambda_A
\lambda_B}$ for the annihilation of $\st + \tilde{t}_1^*$ into $A+B$ with
helicities $\lambda_A$ and $\lambda_B$, respectively. Here, the initial
state is assumed to be a color singlet, and summation over color degrees of
freedom of the final state is understood.\footnote{Following ref.\cite{32} the
color wave function of the initial state in eq.(\ref{e5}) has been normalized
such that the color factor is 1 if $A$ and $B$ are singlets under $SU(3)$;
this explains the factor of 3 in eq.(\ref{e5}). See the Appendix for more
details.} Since we are only interested in $S-$wave initial states we need
the Feynman amplitude only in the limit of vanishing relative velocity $v$ of
the stop squarks; this simplifies the calculation considerably.

We computed the matrix elements for the following processes:
\ben \label{e7} \beq
\sigst &\rightarrow gg \label{e7a}; \\
\sigst &\rightarrow \ww; \label{e7b} \\
\sigst &\rightarrow ZZ; \label{e7c} \\
\sigst &\rightarrow Z \gamma; \label{e7d} \\
\sigst &\rightarrow \gamgam; \label{e7e} \\
\sigst &\rightarrow hh; \label{e7f} \\
\sigst &\rightarrow b \bar{b}; \label{e7g} \\
\sigst &\rightarrow t \bar{t}; \label{e7h} \\
\sigst &\rightarrow \tilde{Z}_i \tilde{Z}_j, i,j=1,\dots,4, \label{e7i}
\eeq \een
where $h$ in eq.(\ref{e7f}) stands for the light neutral scalar Higgs boson.
In general \stst\ annihilation can proceed via the four classes (topologies)
of Feyman diagrams shown in fig.1. The $t-$channel diagram 1a contributes to
all modes of eq.(\ref{e7}), but with different particles being exchanged: \st\
for the $gg, \ Z \gamma$ and \gamgam\ final states; \st\ and \stt\ for the
$ZZ$ and $hh$ final states; $\tilde{b}_L$ for the \ww\ final state; a chargino
for the $b \bar{b}$ final state; a neutralino or gluino for the $t \bar{t}$
final state; and a top quark for the $ \tilde{Z}_i \tilde{Z}_j$ final state.
The $u-$channel diagram of fig.1b only contributes if the final state
particles do not carry any conserved charge (reactions \ref{e7a},c--f,i); the
exchanged particles are then the same as for the $t-$channel diagram. The
4--point diagram of fig.1c only contributes if the final state particles are
bosons (processes \ref{e7a}--f). Finally, in the limit $v \rightarrow 0$ only
the neutral scalar Higgs bosons $h, \ H$ can be exchanged in the $s-$channel
diagram of fig.1d; this diagram therefore only contributes to reactions
(\ref{e7b},c,f--i). Explicit expressions for the matrix elements for the
processes of eq.(\ref{e7}) are listed in the Appendix.

A first example of \sigst\ branching ratios is shown in fig.2. For clarity not
all final states listed in eqs.(\ref{e4}), (\ref{e7}) are represented in this
figure. We have fixed $\mt = -\mu = 150$ GeV, $\mstl = \mstr = 200$ GeV,
$\tanb = 2, \ m_P = 500$ GeV and $M_2 = 100$ GeV, and have varied $A_t$
between --67 and --312 GeV; since $A_t$ and $\mu$ have the same sign,
\msig\ decreases monotonically with increasing $|A_t|$, see
eq.(\ref{e3a}).\footnote{In this and the following figures we have set
$\msig = 2 \mst$, i.e. have neglected the small reduction of \msig\ due to the
binding energy. This is consistent since the treatment of refs.\cite{31,32}
also neglects the binding energy when computing \sigst\ decay widths.}

We see that for this choice of parameters the by far dominant decay mode of
\sigst\ is into two gluons, as long as the single stop decays (\ref{e4a},b)
are kinematically forbidden. In fig.2 the mass of the lighter chargino is
around 110 GeV. For $\msig > 2 \left( m_{\tilde{W}_1} + m_b \right) \simeq
230$ GeV the decay (\ref{e4a}) (not shown) opens up and quickly dominates over
all annihilation decay modes (\ref{e7}). Notice, however, that even for
$\msig = 400$ GeV the width of the single stop decay mode (\ref{e4a}) is
``only" a few hundred times larger than that for the $gg$ final state; since
the partial width for the decay mode (\ref{e4c}) is $\sim 10^{-7}$
smaller than that for (\ref{e4a}) we conclude that this loop decay is
entirely negligible as far as \sigst\ decays are concerned.

Given the large widths for the single stop decay modes (\ref{e4a},b), one
might worry whether our treatment is adequate for these decays. Inherent to
our approach is the assumption that formation and decay of \sigst\ can be
treated separately. This is only true if the \sigst\ formation time is
significantly shorter than its lifetime. A good measure for the (inverse of)
the formation time is the binding energy $E_{bind}$ of \sigst. For a purely
Coulombic potential the time required to complete one Bohr orbit is
proportional to $1/E_{bind}$, and we expect this relation to survive
qualitatively also for the more realistic QCD potential of ref.\cite{33}. In
fig.3 we therefore compare the binding energy of the lowest (1$S$) \sigst\
state, as parametrized in ref.\cite{33}, with the total \sigst\ decay width,
for two different choices of parameters. In both cases we took $\mt =150$ GeV,
$\tanb = 2, \ \mstl = 400$ GeV, $\mstr = 300$ GeV and $m_P = 500$ GeV; \msig\
was varied by changing $A_t$. The solid curve has been obtained with $\mu =
-300$ GeV and fixed $M_2 = 100$ GeV. In this case the light chargino is mostly
an $SU(2)$ gaugino (wino), and the lightest neutralino is mostly a $U(1)$
gaugino (bino). Recall that $\mstl > \mstr$ implies that \st\ is dominantly an
$SU(2)$ singlet ($\cos^2 \theta_t < 1/2$), so that in this scenario the $\st
\tilde{W}_1 b$ coupling $\propto \cos \! \theta_t$ is suppressed; $\cos^2
\theta_t$ decreases with decreasing $\stl - \str$ mixing and increasing \mst\
in this case, which explains the flattening of the solid curve at large \msig.
We see that in this case the total \sigst\ decay width is still a factor 3 to
10 below the binding energy, even well beyond the threshold for decays
(\ref{e4a},b). Our results of fig.2, where $\tilde{W}_1$ is also mostly a
wino, should therefore be at least approximately correct.

In contrast, the dashed curve has been obtained for fixed $\mu = -80$ GeV,
while $M_2$ has been increased along with \mst. The lightest chargino and
neutralino states are now both higgsino--like, so they couple to \stl\ and
\str\ with equal (top Yukawa) strength. The total \sigst\ decay width
therefore increases rapidly with \msig; moreover, the opening of the $\st
\rightarrow t + \tilde{Z}_1$ channel is more pronounced than in the previous
case. As a result, $\Gamma(\sigst)$ does indeed become comparable to the
binding energy in this scenario, which means that our approach will not work
for
$\msig > 300 $ GeV or so. Methods that have recently been developed to
describe the $t \bar{t}$ threshold \cite{34} will have to be adopted
\cite{34a} instead.

Finally, in both cases we observe a very prominent peak at $\msig = 505$ GeV,
where the $s-$channel heavy Higgs boson exchange diagrams become resonant.
Since the total decay width of the heavy Higgs boson exceeds the \sigst\
binding energy, a proper description of this case would have to combine the
methods of refs.\cite{34} with the results of refs.\cite{35} where the mixing
between a nonrelativistic bound state with a (narrow) Higgs resonance is
discussed.

The results of fig.2 show that the branching ratios for those annihilation
decays that might yield a detectable signal for \sigst\ production at hadron
colliders (see sec. 3) become very small if the single stop decays
(\ref{e4a},b) are allowed; fig.3 showed that \sigst\ may not have time to
form at all if there are light higgsino--like states. In fig.4 we have
therefore chosen our parameters such that these single stop decays are
kinematically forbidden for $\msig \leq 600$ GeV; this has been obtained
by choosing $\mu = -300$ GeV, with the other parameters having the same values
as for the dashed curve in fig.3. Comparing fig.4 with fig.2, we notice
two obvious differences. One is the structure around $\msig = 505$ GeV, which
is due to $H$ exchange becoming resonant as already discussed in connection
with fig.3. Of course, in the immediate vicinity of the resonance our results
are not reliable, but the $t \bar{t}$ final state remains dominant in regions
of parameter space where $\Gamma(\sigst)$ is well below the binding
energy.

The other prominent feature of fig.4 is the very rapid increase of the
branching ratios for the $hh$ and, to a lesser extent, \ww\ and $ZZ$ final
states. In case of the $hh$ final state this can be explained from the
observation that in the relevant limit $m_P^2 \gg m_Z^2$ the $h \stst$
coupling \cite{36} contains a term $\propto \mt \left( A_t + \mu \cot \! \beta
\right) / m_W \propto m^2_{LR}/m_W$, where $m^2_{LR}$ is again the
off--diagonal entry of the stop mass matrix (\ref{e1}). Moreover, in fig.4
\mstl\ and \mstr\ are rather large, so that the $hh$ threshold occurs at a
point where $\mstsq \ll \mstlrsq$; eq.(\ref{e3a}) shows that this also
implies $m^2_{LR} \sim \min ( \mstlsq, \mstrsq) \gg \mstsq$. Close to
threshold the \st\ exchange contribution therefore scales like \be \label{e8}
{\cal M}(\stst \rightarrow hh)|_{\tilde{t}_1 \ exchange} \propto
\frac { \min(m^4_{\tilde{t}_L}, m^4_{\tilde{t}_R} )}
{m_W^2 ( \mstsq - m_h^2/2 ) },
\ \ \ \mstlrsq \gg m_t^2, \mstsq, \ee
see eq.(\ref{ap7}); the square of this amplitude clearly decreases very
quickly as we move away from the $hh$ threshold ($\propto
m^{-4}_{\tilde{t}_1}$). This rapid rise of $\Gamma(\sigst \rightarrow hh)$ at
threshold has first been observed in ref.\cite{25}.\footnote{In that paper the
\stt\ exchange contribution to $hh$ production has not been included; this
contribution is small where the $hh$ channel is important.}

The behavior of the matrix elements for the \ww\ and $ZZ$ final states is
somewhat more complicated. In the region $\mstsq \gg m_W^2$ we can use the
equivalence theorem \cite{37} to understand the production of longitudinal
gauge bosons, which in this region usually dominates the production of
transverse gauge bosons. This theorem states that Feynman amplitudes involving
external longitudinal gauge bosons are the same (up to corrections of order
$1/m_W^2$) as those where the gauge bosons are replaced by the corresponding
would--be Goldstone bosons (GBs). These Goldstone modes can also be produced
via the Feynman diagrams of fig.1. However, it is important to note that the
squarks which are exchanged in the $t-$ and $u-$channels are {\em heavy} here.
In case of the neutral GBs only \stt\ contributes since there are no diagonal
couplings between pseudoscalar Higgs bosons and stop squarks \cite{28}. The
corresponding diagrams for charged GBs contain a $\tilde{b}_L$
squark\footnote{Diagrams with $\tilde{b}_R$ exchange are \cite{28}
proportional to $m_b^2$ and can thus be neglected.}; note that \msb\ is
linked to \mstl\ by gauge invariance:
\be \label{e9}
m^2_{\tilde{b}_L} = \mstlsq - m_W^2 \cos \! 2 \beta. \ee
The nonvanishing trilinear scalar couplings of the GBs have similar strength
as the $h \stst$ coupling; the corresponding $t-$ and $u-$channel diagrams
are therefore at best of order \be \label{e10}
{\cal M}(\stst \rightarrow GBs)|_{\tilde{q} \ exchange} \propto
\frac { \min(m^4_{\tilde{t}_L}, m^4_{\tilde{t}_R} )} {m_W^2 \mstlrsq  },
\ \ \ \mstlrsq \gg m_t^2, \mstsq, \ee
i.e. are suppressed by a factor of order $\left( \mst/\mstt \right)^2$
compared to those for the $hh$ final state. In addition, there are $s-$channel
$h$ exchange contributions \be \label{e11}
{\cal M}(\stst \rightarrow hh, GBs)|_{h \ exchange} \propto
\frac { \min(m^2_{\tilde{t}_L}, m^2_{\tilde{t}_R} )}
{ 4 \mstsq - m_h^2}, \ \ \ \mstlrsq \gg m_t^2, \mstsq \ee
This contribution exists for both $hh$ and \ww\ and $ZZ$ final states, but
is suppressed by a factor of order $\left( m_W / \mstlr \right)^2$ compared
to the \st\ exchange contribution (\ref{e8}) to $hh$ production. Far above
threshold the width for longitudinal gauge boson production is therefore
suppressed by a factor of order $\left( \mst / \mstt \right)^4$ compared to
the width for the $hh$ final state.

The equivalence theorem is not applicable close to the \ww\ and $ZZ$
thresholds. Near the thresholds the $t-$ and $u-$channel diagrams, which only
contribute for longitudinal gauge bosons as shown in eq.(\ref{ap3}), are
suppressed by powers of the phase space factor $\beta$ of eq.(\ref{e6}). The
4--point and $s-$channel $h$ exchange diagrams therefore dominate here; the
curves of fig.2 show that they often interfere destructively. Note that for
$\mstsq \ll \mstlrsq$, eq.(\ref{e11}) also applies approximately for the
$s-$channel contribution to \ww\ and $ZZ$ production; it is this term which
leads to the rapid increase of the corresponding partial widths near
threshold.\footnote{The coupling of the heavy scalar Higgs boson $H$ to $W$
and $Z$ bosons is suppressed \cite{28} for $m_P^2 \gg m_W^2$.}

Why did refs. \cite{24} and \cite{26} find so large branching ratios for the
\ww\ final state? The crucial omission is that the relation (\ref{e9}) between
\msb\ and \mstl\ has not been taken into account in these papers. We
emphasize again that this relation follows directly from $SU(2)$ gauge
invariance (and its spontaneous breakdown); it is {\em independent} of the
details of supersymmetry breaking. A violation of eq.(\ref{e9}) therefore
implies explicit (hard) gauge symmetry breaking, which renders the theory
non--unitary and/or non--renormalizable. In addition, \stl--\str\ mixing has
been neglected in these papers. Looking at the exact expression (\ref{ap3}) for
${\cal M}\left( \stst \rightarrow \ww \right)$ it is clear that the
cancellation between the $t-$channel and 4--point diagrams which ensure
unitarity in the simple limit $\cos^2 \tht = 1, \ \mst = \mstl = \msb$ can
be spoiled if one choses $m^2_{\tilde{b}_L} \gg \mstsq$. However, this implies
either $\cos^2 \tht \rightarrow 0$ (if $m^2_{LR}$ is kept fixed), or a large
$s-$channel $h-$exchange contribution which again ``conspires" to restore
unitarity; recall that gauge invariance relates the $h \stst$ coupling to
the \stl\ -- \str\ mass splitting and hence to \msb\ if $\mstl \simeq \mstr$.
In short, the suppression of the partial widths for $\sigst \rightarrow \ww,
ZZ$ is a textbook example for the unitarity restoring cancellations that
are so characteristic for gauge theories.

Figs. 2 and 4 show that the branching ratios for all other modes listed in
eqs.(\ref{e7}) are quite small. The width for the \gamgam\ final state is
simply $8 \alpha^2 / (9 \alpha^2_s) \cdot \Gamma(\sigst \rightarrow gg)$, and
the partial width for the $Z \gamma$ final state is of similar magnitude or
even smaller. (Recall that the $Z \stst$ coupling vanishes \cite{15} for
$\cos^2 \tht = 4/3 \sin^2 \theta_W$.) For the parameter choices of figs. 2
and 4 the light neutralinos $\tilde{Z}_{1,2}$ are gaugino--like; the partial
widths for the $\tilde{Z}_i \tilde{Z}_j$ final states are therefore comparable
to those for the \gamgam\ and $Z \gamma$ final states. Had we chosen the
light neutralinos to be higgsino--like, their partial widths would have been
larger by a factor $\propto \left(\mt/m_W \right)^4$. For parameter choices
leading to mixed--state neutralinos (where both gaugino and higgsino
components are sizable) the $s-$channel $h$ exchange contribution to
$\tilde{Z}_i \tilde{Z}_j$ production can become important, leading to partial
widths comparable to those of the \ww\ and $ZZ$ final states. However, the
existence of light higgsino--like or mixed--state neutralinos in the MSSM
also implies a small mass for the light chargino, so that the single stop
decay mode (\ref{e4a}) is allowed, totally swamping all \sigst\ annihilation
decay modes as we have seen above.

Finally, the partial width for the $b \bar{b}$ final state is very small
unless $\msig \simeq m_h$ or $\tanb \gg 1$. At first glance the gaugino
exchange diagram seems to contribute with full $SU(2)$ gauge strength.
However, chirality implies that $\Gamma(\sigst \rightarrow f \bar{f})
\propto m_f^2$ for any SM fermion $f$. As a result the $b \bar{b}$ final state
can be important only if the $b-$quark Yukawa coupling is enhanced ($\tanb \gg
1$), or in the immediate vicinity of the $h$ resonance. Fig.4 shows that even
the partial width for the $t \bar{t}$ final state is quite small away from the
$H$ pole. This is partly due to destructive interference between $s-$ and
$t-$channel diagrams for $\msig < m_H$, and partly because color factors
suppress all $t-$channel contributions, see eq.(\ref{ap13}).

We thus conclude that, if the tree--level single stop decays (\ref{e4a},b) are
kinematically forbidden and \msig\ is not close to either $m_h$ or $m_H$, the
total \sigst\ decay width is dominated either by the $gg$ or $hh$ partial
width, with the \ww\ and $ZZ$ partial widths playing a secondary role. Our
discussion of fig.4 already showed that the ratio of the $gg$ to $hh$ partial
widths crucially depends on the size of the $LR$ element of the stop mass
matrix (\ref{e1}). This is further illustrated in fig.5, where we show the
$gg, \ hh$ and \ww\ partial widths as a function of \msig\ for three different
choices of parameters. We have fixed $\mu = 500$ GeV, $m_P = 1$ TeV and $\mt =
150$ GeV; instead of varying $A_t$ we have fixed the dimensionless quantity $A
\equiv 2 A_t / (\mstl + \mstr)$ as well as the ratio $\mstr/\mstl$, and varied
\mstl. The dependence of the $hh$ and \ww\ partial widths on \msig\ is
therefore quite different than in figs. 2 and 4.\footnote{Of course, the $gg$
partial width is fixed uniquely by \msig\ and the strength of the QCD coupling
constant, and his hence the same for all three cases.} In particular,
eq.(\ref{e3a}) implies that now $m_{LR}^2$, and hence the strength of the $h
\stst$ coupling, increases with increasing \msig\ even if $\mstsq \ll
\mstlrsq$. For large $A$ the $hh$ partial width still decreases with
increasing \msig, due the \st\ propagator suppression, but the decline is much
less rapid than in fig.4. Notice that we have chosen $\mu> 0$ here, so that
$m^2_{LR}$ increases monotonically with increasing $A$. However, due to
destructive interference between different diagrams both the $hh$ and the \ww\
partial width initially decrease with increasing $m^2_{LR}$, shooting up
quickly once $A>1$. We have already seen above that the partial width for the
\ww\ final state always stays well below those for the $hh$ and $gg$ final
states. Here we see that the off--diagonal entries of the stop mass matrix
(\ref{e1}) need not be all that large for the $hh$ mode to dominate \sigst\
decays. Finally, the short dashed curve has been obtained with $\mstl =
\mstr$, as compared to $\mstr = 0.7 \ \mstl$ for the other curves; we see that
this has only little effect on $\Gamma(\sigst \rightarrow hh)$. Since we kept
$A$ fixed, the size of $m^2_{LR}$ for given \mst\ is about the same for the
two choices of $\mstr/\mstl$; this again indicates that the size of $m^2_{LR}$
is indeed the quantity that decides whether or not the $hh$ partial width is
sizable. We will come back to this point later.

Having completed our discussion of \sigst\ decays, we are now ready to study
possible signals for its production at hadron colliders.
\setcounter{footnote}{0}
\section*{3) Signals for stoponium production}
In this section we discuss how one might search for \sigst\ states. We focus
on hadron (super)colliders, since they offer the largest cross sections;
furthermore, the machines we discuss either already exist (tevatron) or are in
a relatively advanced stage of planning (LHC), while plans for future linear
$e^+e^-$ or \gamgam\ colliders are still at a very preliminary stage. The
production of \sigst\ at hadron colliders proceeds via gluon fusion. This
process is related by crossing to the $\sigst \rightarrow gg$ decay, whose
partial width we computed in the previous section. The total cross section
for \sigst\ production is then (to leading order in the QCD coupling constant)
simply given by \be \label{e12}
\sigma(pp \rightarrow \sigst + X) = \int_{\tau}^1 \frac{dx}{x} \tau
G(x,Q^2) G \left( \frac{\tau}{x}, Q^2 \right) \frac {\pi^2} {8
m^3_{\sigma_{\tilde t}} } \Gamma ( \sigst \rightarrow gg), \ee
where $\tau \equiv m^2_{\sigma_{\tilde t}}/s$ with $s$ being the squared $pp$
invariant energy; since in leading order only gluon fusion contributes, the
cross section is the same for $pp$ and $p \bar{p}$ colliders. Unless stated
otherwise we have used the parametrization of ref.\cite{46} for the gluon
distribution functions $G$, and have chosen the scale $Q=\mst$. In order
to set the stage for the subsequent discussion we remark here that
eq.(\ref{e12}) predicts a total \sigst\ production cross section at the LHC
($\sqrt{s}=14$ TeV) of 80 (22, 8.5) pb for \msig=150 (200, 250) GeV.

Unfortunately most \sigst\ decays will not lead to a signal that is detectable
at hadron colliders. To begin with, the QCD di--jet cross section integrated
over any reasonable invariant mass window will be many orders of magnitude
larger than the total \sigst\ production cross section, making it impossible
to detect $\sigst \rightarrow gg$ decays. QCD backgrounds also swamp $\sigst
\rightarrow b \bar{b}, \ t \bar{t}$ decays. Single stop $\st \rightarrow b +
\tilde{W}_1$ decays can give hard, isolated leptons in the final state if
$\tilde{W}_1$ decays leptonically. However, the open \stst\ pair production
cross section \cite{14} is some 4 orders of magnitude larger than the \sigst\
cross section. The presence of 2 invisible LSPs and at least one neutrino in
the $\sigst \rightarrow \tilde{W}^+_1 \tilde{W}^-_1 b \bar{b} \rightarrow
l^{\pm} X$ signal makes it impossible to reconstruct \msig\ even in principle.
We conclude that single stop decays will never give a signal for \sigst\
production at hadron colliders even if conditions are favorable for the
detection of open stop production \cite{14}.

In ref.\cite{26} the use of the \ww\ decay mode (for rather heavy \sigst, to
be produced at future supercolliders) was advocated. However, we have seen in
the previous section that in this paper the $Br(\sigst \rightarrow \ww)$ was
overestimated by a large factor. Moreover, this final state can only compete
with QCD backgrounds if both $W$ bosons decay leptonically. The event will
then contain two neutrinos, which make it impossible to reconstruct the
invariant mass of the \ww\ system. Even if it were possible to somehow
discriminate against the enormous $t \bar{t}$ background at hadron
supercolliders, the continuum cross section for \ww\ production is still
at least an order of magnitude larger \cite{37a} than the $\sigst \rightarrow
\ww$ signal.

The authors of ref.\cite{25} proposed to use the $\sigst \rightarrow hh
\rightarrow \fourtau$ decay as a signal. This might have been feasible for
light \sigst\ and light $h$ at the tevatron collider. Unfortunately the lower
bound on $m_h$ from Higgs searches at LEP \cite{38} excludes this possibility
for \sigst\ light enough to be produced at the tevatron. The \fourtau\ SM
background is much larger at supercollider energies, making it considerably
more difficult to observe a \sigst\ signal in this channel. For example, at
the LHC ($\sqrt{s} = 14$ TeV) the $ZZ \rightarrow 4 \tau$ background amounts
\cite{37a} to approximately 10 fb. Using \cite{39} $Br(h \rightarrow \tau^+
\tau^-) = 10\%$ we find that the $\sigst \rightarrow 4 \tau$ signal could be
as large as 1 pb if $\msig \leq 150$ GeV and $Br(\sigst \rightarrow hh) \simeq
1$. While this is considerably larger than the most narrowly defined physics
background, it is smaller than the cross section for $pp \rightarrow b b
\bar{b} \bar{b} \rightarrow 4 \tau$ production\footnote{This is true for the
high luminosity option of the LHC, where most $4 b$ events originate from
independent $pp$ collisions. The $4b \rightarrow 4\tau$ background for low
luminosities is\cite{39a} around 0.2 pb.}. Backgrounds from events containing
jets with low charged particle multiplicity, which might fake $\tau$ signals,
could be even more dangerous. Note that the presence of (at least four)
neutrinos in the final state makes it once again impossible to reconstruct
either $m_h$ or \msig. Isolation of the \fourtau\ signal at a hadron
supercollider therefore seems quite difficult to us; certainly detailed Monte
Carlo simulations would have to be performed before a good case for this
signal can be made.

This leaves us with \sigst\ decay modes with rather small branching ratios.
The $\tilde{Z}_i \tilde{Z}_j$ mode again suffers from the problem that \msig\
cannot be reconstructed since the final state contains two LSPs. The $ZZ$
mode offers a clean signal if both $Z$ bosons decay leptonically.
Unfortunately the branching ratio for the $ZZ$ final state is often quite
small, as shown in figs. 2 and 4. The conditions for this signal are most
favorable for large $LR$ element of the stop mass matrix (\ref{e1}) and in the
mass range $m_h \geq \mst > m_Z$. In this case $Br(\sigst \rightarrow ZZ)$ can
be as large as 10\%, giving a maximal $\sigma (pp \rightarrow \sigst
\rightarrow \fourl X) \simeq 8 fb$ for $\msig = 200$ GeV at $\sqrt{s} = 14$
TeV, corresponding to 800 events in a full LHC year ($\int {\cal L} dt = 100$
fb$^{-1}$); this should be readily detectable. However, since in the MSSM
$m_h$ cannot be larger than 140 GeV or so even after the inclusion of
radiative corrections \cite{29,30} this window of opportunity is rather
narrow. Figs. 2 and 4 show that more generically the $ZZ$ branching ratio lies
between 0.1 and 1\%, making this signal rather marginal; recall that in the
SM, $\sigma (pp \rightarrow ZZ \rightarrow \fourl X ) \simeq 40$ fb \cite{37a}
at the LHC.

The $Z \gamma$ final state could also give a clean signal if $Z \rightarrow
l^+l^-$. Unfortunately the combined branching ratio for $\sigst \rightarrow
Z \gamma \rightarrow l^+l^- \gamma$ is always below 0.01\%. Note that
photons have to be quite energetic to yield a potential signal at the LHC;
this final state can therefore only be used for $\msig > 150$ GeV or so.
The total $pp \rightarrow \sigst \rightarrow Z \gamma \rightarrow l^+l^-
\gamma$ signal then amounts to at most 10 fb at $\sqrt{s}=14$ TeV; in
comparison, the SM physics background \cite{40} is about 600 fb even if one
requires the transverse momentum of the photon to exceed 50 GeV. It seems
therefore very unlikely to us that this signal will be detectable.

Such considerations led us to propose \cite{27} the \gamgam\ final state as
the most promising signal for \sigst\ production at hadron colliders. Figs. 2
and 4 show that the corresponding branching ratio is typically a few times
$10^{-3}$, although it can be substantially smaller near an $s-$channel pole
or for large $m^2_{LR}$; this is considerably larger than typical branching
ratios into leptonically decaying $Z$ bosons. Since the $\sigst \rightarrow
\gamgam$ partial width is determined uniquely by \mst\ (for given QCD
potential), the \gamgam\ signal rate depends on model parameters only via the
total \sigst\ decay width. The signal is very simple, consisting of two hard
photons with invariant mass $\mgg=\msig$ in a hadronically quiet event. Of
course, there is also a sizable SM background from $q \bar{q}$ annihilation
and $gg$ fusion. It has been studied in some detail in the literature
\cite{41,42} as a background to a possible signal for intermediate mass Higgs
boson production. Recall that the natural width of the signal peak in our case
is just a few MeV, see fig.5; in contrast, the background gives a smooth
distribution in \mgg. The question is then if, or under what circumstances,
the signal peak is observable on top of the background.

In most SM \gamgam\ events the photons will emerge at small angles, due to
$t-$ and $u-$channel quark propagator effects; in contrast the signal is
isotropic in $\cos \! \theta^*$, where $\theta^*$ is the scattering angle in
the \gamgam\ center--of--mass system. We therefore impose the cut
\be \label{e13}
| \cos \! \theta^* | \leq 0.5. \ee
The \gamgam\ background has been computed to next--to--leading order in QCD
\cite{42}. However, if one vetoes against the presence of hard, central jets
in the event and requires the photons to be isolated, the NLO prediction for
the background rate is actually very similar to the leading order estimate.
Moreover, no NLO calculation for the signal cross section exists as yet. We
therefore also treat the background in leading order, but we include the
$gg \rightarrow \gamgam$ contribution which has been found to be very
important \cite{41} especially for low \mgg.

As noted above, the natural width of the signal peak is extremely small;
however, due to detector resolution effects its actual (measured) width will
be much larger. Clearly the background should be integrated over this
larger range of \mgg. On the other hand, the signal within a given bin need
not be larger than the background in order to be detectable, since the
expected background level can be determined experimentally by fitting a smooth
function to the sidebins. The question is then whether the excess in the
signal bin is statistically significant or not. Following ref.\cite{43}, where
the search for SUSY Higgs bosons was discussed, we define the signal to be
significant if the 99\% c.l. upper limit on the background rate is smaller
than the 99\% c.l. lower limit on signal plus background combined. In the
limit of large event numbers, where Gaussian statistics can be used, this
means: \be \label{e14}
N_b + N_s - 2.32 \sqrt{N_b + N_s} \geq N_b + 2.32 \sqrt{N_b}, \ee
which implies \be \label{e15}
N_s \geq 2.32 \left( 2 \sqrt{N_b} + 2.32 \right). \ee
Here, $N_b$ and $N_s$ are the expected number of signal and background
events after cuts. As noted earlier, the background has to be integrated over
a detector--dependent bin width $\Delta \mgg$: \be \label{e16}
N_b = \left( \int {\cal L} dt \right) \left. \cdot \frac {d \sigma_b}
{d \mgg} \right|_{\mgg=\msig} \cdot \Delta\mgg. \ee
In the limit $\sqrt{N_b} \gg 1$ the minimal detectable signal cross section
$\sigmin = N_s/ \left( \int {\cal L} dt \right)$ therefore
scales like the inverse square root of the integrated luminosity and also like
the inverse square root of the energy resolution of the electromagnetic
calorimeter, which determines the size of $\Delta \mgg$. In our background
estimates we simply took $\Delta \mgg$ to be twice the assumed invariant mass
resolution.

In fig.6 we show the expected signal at the tevatron ($\sqrt{s}=1.8$ TeV). In
addition to the cut (\ref{e13}) we have required that both photons have
rapidity $|y_{\gamma}| \leq 1.1$, so that $|\cos \! \theta_{\gamma}| \leq 0.8$
in the lab frame; the same cut has been applied by the CDF collaboration in
their preliminary analysis \cite{44} of events with two hard photons. The
dashed curve has been obtained under the assumption that the total width of
\sigst\ is determined by the $gg$ and \gamgam\ partial widths alone, while for
the solid curve all \sigst\ decay modes of eq.(\ref{e7}) have been included.
The two results are indistinguishable except for $\msig \simeq m_h =87$ GeV
for the given choice of parameters. The signal for such light \sigst\ does
therefore not depend on the details of the (s)particle spectrum (aside from
\mst) as long as \st\ has no tree--level two--body decays and $s-$channel
$h-$exchange contributions to \sigst\ decays are not ``accidentally''
enhanced.

In fig.6 we also show our estimates for the minimal detectable signal cross
section (dotted lines) for three different values of the integrated luminosity
representing the present status (18 pb$^{-1}$), the hoped--for luminosity
after run Ib (100 pb$^{-1}$), and an estimate of what might be achievable
after the new Main Injector has been completed (1 fb$^{-1}$). Here we have
assumed an invariant mass resolution of 2\%, i.e. $\Delta \mgg = 0.04 \mgg$.
Since the expected number of background events per bin is not always large
we have used Poissonian statistics to derive these curves; however,
eq.(\ref{e15}) gives quite similar results in the mass range where detection
of \sigst\ might be possible.

We conclude from fig.6 that the mass reach of the tevatron for \sigst\
searches is quite modest. It is therefore exceedingly unlikely that one of the
two CDF events \cite{44} with very large \mgg\ (350 and 430 GeV, respectively)
is due to \sigst\ production. Indeed, most of the region that one might be
able to probe even with an integrated luminosity of 1 fb$^{-1}$ is already
excluded by LEP measurements of the total $Z$ decay width and/or squark
searches at LEP, {\em unless} the $Z \stst$ coupling happens to be suppressed
by $\stl-\str$ mixing. On the other hand, fig.6 also reveals that {\em
existing} tevatron data might already help to probe this light stop window; in
particular, they might confirm or rule out the recently proposed explanation
\cite{eno} of the (small) excess of events containing low$-p_T \ D^*$ mesons
observed \cite{topaz} by the TOPAZ collaboration in terms of \stst\ production
and subsequent $\st \rightarrow c + \tilde{Z}_1$ decay with $\mst \simeq 15$
GeV and $m_{\tilde{Z}_1} \simeq 12.5$ GeV.

In fig.7 we show results for LHC energy ($\sqrt{s}=14$ TeV). We have again
applied the cut (\ref{e13}) on $\cos \! \theta^*$, but have relaxed the
requirement for $y_{\gamma}$ slightly compared to fig.6; our cut $|y_{\gamma}|
\leq 1.74$ still ensures that the photons are at least 20$^{\circ}$ away from
the beam pipes, i.e. are well isolated from the beam remnant jets. We have
also rather optimistically assumed a 1\% resolution for the measurement of
\mgg, i.e. used eq.(\ref{e16}) with $\Delta \mgg = 0.02 \mgg$. Since now the
expected number of background events per bin is quite large we have used
Gaussian statistics to estimate the minimal detectable signal \sigmin; the
dotted curve shown in fig.7 assumes one nominal LHC year of operations, i.e.
$\int {\cal L} dt = 100$ fb$^{-1}$.

The solid line in fig.7 shows the total \sigst\ production cross section
(divided by 100), without any cuts. The dashed curves show the \gamgam\
signal cross sections after cuts for the two sets of parameters chosen in
figs. 2 and 4. We saw in sec. 2 that the branching ratio for the \gamgam\
mode is about $3 \cdot 10^{-3}$ at small \msig, where the $gg$ mode dominates.
Comparing the solid and dashed lines in fig.7 we see that our cuts reduce
the signal by approximately a factor of 3.5 at low \msig; at high \msig\
almost half of all $\sigst \rightarrow \gamgam$ events pass. Of course, the
cut (\ref{e13}) alone excludes 50\% of all signal events; for large \msig\
the events are automatically central so that the cut on the rapidity does not
reduce the event number further. The reduction of the $q \bar{q}, \ gg
\rightarrow \gamgam$ backgrounds by our cuts is much larger; in addition we
have to require the photons to be well isolated from all jet activity,
including the beam remnant jets, in order to suppress the bremsstrahlung
background, which otherwise dominates \cite{41}.

The short dashed curve shows that, as anticipated, the \gamgam\ signal
quickly becomes unobservable once \st\ has tree--level two--body decays
($\msig > 230$ GeV; see fig.2). Here the situation is quite analogous to the
case of $t \bar{t}$ bound state production where the \gamgam\ signal also
becomes inaccessible \cite{45} once $m_t > 120$ GeV or so. The rapid decline
of the long dashed curve in fig.7 at $\msig = 180$ GeV is due to the
opening of the $hh$ mode, see fig.4. In this case the \gamgam\ signal
becomes marginal just beyond the $hh$ threshold, but should still be observable
after several years of LHC operations if our assumptions about the detector
resolution can be realized. Note that the signal actually increases with
increasing \msig\ as we leave the $hh$ threshold region, in spite of the
rapid decrease of the total cross section for \sigst\ production; this once
again illustrates the steep decline of the $hh$ partial width with decreasing
$m^2_{LR}$, which corresponds to increasing \mst\ in this case as discussed
in sec. 2. Finally, for $\msig \simeq m_H$ the \gamgam\ signal again becomes
unobservable, due to the large $s-$channel enhancement of the $t \bar{t}$
partial width.

Before we try to further evaluate the stoponium discovery potential of the
LHC it might be worthwhile to discuss some of the uncertainties inherent to
our calculation of signal rates. As stated earlier, the cross sections shown
in figs. 6 and 7 have been computed using the parametrization of the parton
distribution functions in the proton given by Owens \cite{46}; we found that
other recent parametrizations \cite{47} give very similar results. Our cut on
the rapidity of the photon ensures that we probe the gluon density $G$ at
comparatively large values of $x$ where differences between existing
parametrizations are not so large. In our previous figures we have taken
$Q=\mst$ for the momentum scale in the gluon distribution function; the same
choice has been used for the solid and dashed curves in fig.8. In contrast,
the dotted curve has been obtained with with $Q=\msig$. Clearly this change of
$Q$ by a factor of 2 has little effect on the predicted signal rate. For
small (large) values of $x$ the gluon density increases (decreases) as $Q$ is
increased; most of our signal comes from the cross--over region in between,
where $G$ depends very weakly on $Q$. We note here that we have not changed
the scale in \as\ in fig.8, which determines $\Gamma(\sigst \rightarrow gg)$
and hence the total \sigst\ production cross section, see eq.(\ref{e12}).
However, for the given choice of parameters the total \sigst\ decay width is
always dominated by the $gg$ partial width, so that the branching ratio for
the \gamgam\ final state is inversely proportional to $\Gamma(\sigst
\rightarrow gg)$; the signal is therefore almost independent of the choice
of the momentum scale to be used in \as\ in the given case.

Nevertheless our results do depend on the choice of the QCD scale $\Lambda$,
as also shown in fig.8. The reason is that larger values of $\Lambda$
imply a bigger QCD coupling constant \as, and hence a more tightly bound
stoponium system, i.e. larger $|R(0)|^2$; note that the signal is $\propto
|R(0)|^2$ if the total \sigst\ decay width is dominated by annihilation
decays. Ref.\cite{33} provides parametrizations of this quantity for four
different values of $\Lambda$; our previous results have been obtained with
$\Lambda = 0.2$ GeV, which is in between the extreme choices of 0.1 and 0.4
GeV. We see that even for $\Gamma_{\rm tot}(\sigst) \simeq \Gamma(\sigst
\rightarrow gg)$ the variation of $\Lambda$ corresponds to a 30\% uncertainty
of our signal. This uncertainty is even larger if $\Gamma_{\rm tot}(\sigst)
\gg \Gamma(\sigst \rightarrow gg)$. If the total width is dominated by
annihilation decays into $hh$ or $t \bar{t}$ final states, the uncertainty
in $\Lambda$ leads to an approximately 50\% uncertainty of the signal, since
now the increase of the $gg$ partial width, i.e. of the total cross section
for \sigst\ production, is no longer cancelled by a corresponding decrease of
the branching ratio for the \gamgam\ final state when $\Lambda$ is increased.
The $\Lambda$ dependence becomes stronger yet if tree--level two--body decays
of \st\ are possible, since in this case the signal is $\propto \Gamma(\sigst
\rightarrow gg) \Gamma(\sigst \rightarrow \gamgam) / \Gamma_{\rm tot} (
\sigst) \propto \alpha^2_s |R(0)|^4$; the signal now increases by more than
a factor of four when $\Lambda$ is increased from 0.1 to 0.4 GeV. A similarly
strong dependence on $\Lambda$ was observed in ref.\cite{45} for the analogous
case of the \gamgam\ signal for toponium production. However, fig.7 shows
that detection of \sigst\ at the LHC becomes much more difficult if
$\Gamma_{\rm tot}(\sigst) \gg \Gamma(\sigst \rightarrow gg)$, and all but
impossible if \st\ has unsuppressed tree--level two--body decays. The
situation depicted in fig.8 is therefore more characteristic for situations
where the discovery of \sigst\ seems feasible at the LHC.

Yet another uncertainty comes from the existence of higher (excited) stoponium
states. So far, we have only considered the direct production of the lowest
lying ($n=1$) state. However, already for the $c \bar{c}$ system two ($J=1$)
$S-$wave bound states are known to exist; there are three $J=1 \ s-$wave
$b \bar{b}$ bound states. For a Coulomb potential the number of bound states
increases proportional to the square root of the mass of the heavy (s)quark.
As mentioned earlier, in ref.\cite{33} the mass (binding energy) and wave
function at the origin of the first ten heavy (s)quarkonium states are given
(the ground state and nine excited states). Not all of these states will be
true bound states; some may be resonances that decay rapidly into a pair of
stop--flavored (spin--1/2) ``mesons". We assume, rather conservatively, that
the mass of the lightest stop ``meson" lies just 200 MeV above \mst, i.e.
we treat the $n-$th \stst\ state $\sigst(n)$ as a bound state if
\be \label{e17}
m_{\sigma_{\tilde{t}}(n)} < 2 \mst + 0.4 \ {\rm GeV}. \ee

Of course, the wave function at the origin, and hence the production cross
section, is smaller for excited states than for the ground state ($|R(0)|^2
\propto 1/n^2$ for a Coulomb potential). Nevertheless fig.9 shows that the
enhancement of the signal due to the production of excited states can be quite
substantial. In this figure we show $\sum_n |R_n(0)|^2 / |R_1(0)|^2$, where the
sum runs over all states that satisfy the condition (\ref{e17}), i.e. are
true bound states. This ratio is equal to the enhancement of the signal due
to the production of excited stoponium states if annihilation decays dominate
the total \sigst\ decay width, and if the excited states themselves contribute
to the signal in the same way as the ground state does.

This second requirement deserves a brief discussion. Since all excited states
in the sum are $S-$wave states, expression (\ref{e5}) also describes their
annihilation decays (with $R(0)$ replaced by $R_n(0)$). The various stoponium
states should lie close enough together that their \gamgam\ invariant mass
peaks will not be distinguished from each other by the detector, unless we
have underestimated the resolution to be achieved by future experiments.
Annihilation decays of the excited states will therefore contribute to the
signal in the same was as for the ground state. However, these excited states
have additional decay channels: They can decay into lower lying stoponium
states, plus a photon or a mesonic system with vanishing charge and strong
isospin. Cascade decays of excited states into lower lying $S-$wave states
will also contribute to the signal if the lower lying state decays into two
photons, since the existence of additional very soft photons or mesons from
the cascade will hardly be detectable at hadron colliders.

However, an excited state can also decay into a lower lying state with
{\em different} angular momentum, e.g. into a $P-$wave state. The relative
branching ratios for annihilation decays of these higher spin states can
differ significantly from those of the $S-$wave states; eq.(\ref{e5}) is
no longer applicable here. In particular, $P-$wave states do not contribute
to the signal at all (unless they in turn decay into another $S-$wave state),
since they cannot decay into two photons. In order to estimate how much the
higher $S-$wave states contribute to the signal one would therefore have to
follow all their decay chains; this necessitates a complete understanding of
stoponium spectroscopy, which is well beyond the scope of this paper. We
mention here that only a rather small fraction of $\Upsilon(2s)$ and
$\Upsilon(3s)$ mesons decays into $P-$wave $b \bar{b}$ ($\chi_b$) states
which do not decay back into $S-$wave states ($\sim 13\%$ for $\Upsilon(2s)$
and $\sim 22\%$ for $\Upsilon(3s)$). If this result carries over to the
stoponium system, fig.9 should give a good estimate for the enhancement of the
signal due to the production of excited states. Finally, there is a
contribution to the signal from the direct production of states with
$J \ne 0$ and their subsequent decay into $S-$wave states, but it
should be quite small.\footnote{In the most simple treatment of
nonrelativistic bound states\cite{31,32}, the production cross section of
higher spin states is predicted to be small since it is proportional to the
square of derivatives of the wave function at the origin, divided by
additional powers of \mst. Recently Bodwin et al. \cite{bra} have suggested
that the production of $P-$wave quarkonia states might be enhanced by the
presence of a sizable component of the wave function where the $Q \bar{Q}$ is
in a color octet state. However, the same component would also suppress the
branching ratio for $P-$wave states to decay into $S-$ wave states, so that
altogether the $P-$wave contribution to the \gamgam\ signal is still small.}

 In any case, in the absence of a more reliable
treatment of the decays of excited squarkonium states we have conservatively
decided to only include the direct production of the lowest ($n=1$) $S-$wave
state in our estimates of signal cross sections.

We finally address the question of the \sigst\ discovery potential of the LHC,
using the \gamgam\ decay mode and our conservative estimate of the signal
cross section. We have already stated repeatedly that \sigst\ will be
unobservable at hadron colliders, and may indeed not form at all, if the
single stop decay modes (\ref{e4a},b) are unsuppressed {\em or} \sigst\ is
very close in mass to one of the two scalar Higgs bosons of the MSSM; for the
subsequent discussion we therefore assume that this is not the case. We saw in
sec. 2 that the size of the \gamgam\ branching ratio is then almost uniquely
determined by the partial width for the $hh$ final state, the ratio of the
$gg$ and \gamgam\ partial widths being fixed by QCD. Moreover, we saw in fig.
5 that, at least for parameters where $\Gamma(\sigst \rightarrow hh) \geq
\Gamma(\sigst \rightarrow gg)$, the partial width for the $hh$ final state is
determined by the size of the $LR$ element of the stop mass matrix (\ref{e1}).
Under the given assumptions the detectability of the $\sigst \rightarrow
\gamgam$ signal at the LHC therefore basically depends on two parameters: The
mass \msig, which determines the total \sigst\ production cross section; and
$m^2_{LR} = -m_t (\amu)$, which determines the size of the branching ratio of
the \gamgam\ decay mode.

It can safely be assumed that the mass of the top quark will be known quite
precisely before LHC experiments are ready to search for \sigst\ production.
In fig.10 we therefore show the region in the plane of \msig\ and \amu\
that can be probed after one and five nominal LHC years (${\cal L} = 100$
fb$^{-1}$ per year). Here we have assumed $m_t = 150$ GeV and $\tanb=2$, but
this choice has little affect on the accessible region.\footnote{It does
affect the size of the region in the top left corner where the LEP Higgs bound
\cite{38} is violated.}

The general shape of the curves is easy to understand. At small \msig\ the
$hh$ mode is only open if \amu\ is very large; note that radiative corrections
reduce $m_h$ if $\amu \gg \mstlr$ \cite{29,30}. In this case the $hh$ partial
width is very large just beyond the threshold, as shown in fig.4, and the
\gamgam\ signal remains unobservable even after a long running period. On the
other hand, for larger \msig\ the $hh$ channel is always open. We see that the
curves for the maximal accessible \amu\ become quite flat in this region. The
reason is that increasing \msig\ decreases the total cross section for \sigst\
production, but at the same time decreases the branching ratio for the $hh$
mode if \amu\ is kept fixed, see eq.(\ref{e8}). Moreover, the minimal
detectable signal cross section decreases with increasing \msig, although more
slowly than the total \sigst\ production cross section does, as shown in fig.
7. These effects tend to cancel each other, leading to the observed
flattening of the curves for $\msig \geq 220$ GeV. Eventually, however,
increasing \msig\ reduces the $hh$ partial width to a value below the $gg$
partial width; decreasing it even further then has little effect on the
signal, and the curves terminate rather abruptly. We finally note that the
little bulge in the accessible regions at $\amu \simeq 600$ GeV occurs
because for moderate values of $m_{LR}^2$ the $hh$ partial width no longer
grows monotonically with \amu, as we already saw in fig.5.
\setcounter{footnote}{0}
\section*{4) Summary and Conclusions}
In this paper we have studied the decays of $S-$wave \stst\ bound states
\sigst, as well as possible signals for their production at hadron colliders.
We first argued in sec. 1 that there are no strict bounds on \mst\ which hold
both for all $\stl-\str$ mixing angles \tht\ and all values of the LSP mass;
even under relatively mild assumptions a \st\ as light as 40 GeV is still
allowed. This leaves a wide mass region to be explored. We have seen that
\sigst\ production is only detectable at hadron colliders if \st\ has no
unsuppressed tree--level two--body decays. Otherwise single squark decays
of \sigst\ dominate over annihilation decays, and stoponium production gives at
best a small contribution to the signals for open stop production.

The dominant annihilation decay modes of \sigst\ are those into two gluons,
two light scalar Higgs bosons $h$, or a $t \bar{t}$ pair. Since the latter two
decays involve electroweak rather than strong couplings, their partial widths
have to be enhanced dynamically in order to be comparable to or larger than
the one for the $gg$ final state. In case of the $t \bar{t}$ mode this can
only happen if \msig\ is very close to the mass of the heavy scalar Higgs
boson $H$, so that $s-$channel $H-$exchange contributions become (almost)
resonant. The $hh$ partial width becomes large if the off--diagonal entry
$m^2_{LR}$ of the stop mass matrix is approximately as large as the diagonal
entries of that matrix; in such a situation mixing greatly reduces the mass
of the lighter stop eigenstate. Since the $h \stst$ coupling increases with
$m^2_{LR}$ while \mst\ decreases, thereby further enhancing \st\ exchange
diagrams, the $hh$ partial width is very sensitive to $m^2_{LR}$, as
illustrated in fig.5.

Unfortunately we saw in sec. 3 that none of these three potentially dominant
final states leads to a readily detectable signal at hadron colliders. The
most promising mode appears to be the $\sigst \rightarrow \gamgam$ decay,
which gives rise to a peak in the two--photon invariant mass spectrum. We
analyzed this signal in some detail, comparing it to the \gamgam\ continuum
background. We found that existing Tevatron data might already begin to close
the light stop window left by LEP (where the $Z \stst$ coupling is suppressed
by mixing and the \st--LSP mass difference is small). On the other hand, even
fo
r $\int {\cal L}
dt = 1$ fb$^{-1}$ the mass reach of the tevatron only extends to $\msig=90$
GeV. Under favorable circumstances this mass reach can be extended to 500
(700) GeV after one (five) year(s) of operation at the LHC with full
luminosity (${\cal L} = 100$ fb$^{-1}$ per year). Recall, however, that for
$\msig > 120$ GeV the $hh$ decay mode of \sigst\ might be open, which might
greatly reduce the branching ratio for the \gamgam\ final state. More
generally LHC experiments will therefore only be able to probe a region in
the $(\msig, m^2_{LR})$ plane, see fig.10.

We should remind the reader here that our calculation has considerable
uncertainties, even beyond those intrinsic to any leading order QCD prediction
for hadronic processes. On the one hand, we have ignored backgrounds from
jets with very few charged particles, which could fake a single photon. This
background is clearly detector dependent, but could potentially be sizable.
On the other hand, our estimate for the signal rate is probably also too low,
since we have ignored all contributions involving higher (excited) stoponium
states. We saw in fig.9 that they might enhance the signal by as much as a
factor of two; however, a quantitative treatment of their contribution
requires a detailed understanding of the entire stoponium system.

Once \sigst\ production has been observed in the \gamgam\ channel its mass
will be known precisely. If 160 GeV $\leq \msig \leq$ 300 GeV one might then
be able to find evidence of \sigst\ production also using the $\gamma Z$
and/or $ZZ$ channel, where $Z$ bosons decay into $e^+e^-$ or $\mu^+\mu^-$
pairs; at least this task should be easier than searching for \sigst\
production in these channels before \msig\ is known. Once the existence of
\sigst\ has been established one might also try to look for its $hh$ decay via
the \fourtau\ final state. We saw that the cross section for this final state
could be as large as 1 pb; the main problem here is to cleanly identify the
$\tau$ leptons. Data taken at lower luminosity are probably more useful for
this purpose, since the presence of multiple overlapping events will make
$\tau$ identification even more difficult.

Once \msig\ is known, one can even contemplate studying it in some detail at
a \gamgam\ collider. At least in principle such a device can be constructed
\cite{48} by backscattering laser photons off the electrons and positrons of
an $e^+e^-$ collider. The cross section for \sigst\ production could be of
the order of (0.5 pb)/$(\msig/100\ {\rm GeV})^3$. Moreover, by polarizing the
incident photons one can greatly reduce backgrounds; e.g., $\gamgam
\rightarrow q \bar{q}$ production would be suppressed for light quarks if both
photons have the same polarization, which might even allow to detect $\sigst
\rightarrow gg$ decays. The strong dependence of many partial widths on model
parameters (see figs.2 and 4) makes their measurement either at a $pp$ or a
\gamgam\ collider very interesting, and in particular offers one of the few
possibilities to measure the size of the trilinear soft breaking parameter
$A_t$.

Notice that searches for stoponium production are in some sense complementary
to searches for open stop production. Stoponium states will be very difficult
to detect, and might not form at all, if \st\ decays via two--body modes that
are accessible at tree level. On the other hand, open stop production at
hadron colliders will be difficult to detect either via its semi--leptonic
decay or via a missing $p_T$ signal unless the \st--LSP mass difference is
sizable. These two requirements are complementary because within the minimal
SUSY model there is a strong correlation between the possibility of
tree--level two--body decays of \st\ and a large \st--LSP mass difference.
This is obvious for the stop $\rightarrow$ top $+$ LSP decay, but also holds
if the $\st \rightarrow b\ +$ chargino decay is allowed, at least in the case
where the LSP (which we always assume to be the lightest neutralino) is
dominantly a gaugino. A gaugino--like LSP is favored dynamically in models
with radiative gauge symmetry breaking \cite{10}, as well as by cosmological
considerations; unlike a higgsino--like or mixed--state LSP, it can naturally
explain the observed Dark Matter in the Universe \cite{49}.

We thus conclude that there should be a sizable \st--LSP mass difference,
which facilitates detection of open stop production, if \st\ decays rapidly;
if the \st--LSP mass difference is small, the light stop is usually long lived
and chances for stoponium production should be good. This complementarity is
not perfect. On the one hand, the possibility of a large branching ratio of
\sigst\ into $hh$ or, worse, $t \bar{t}$ final states means that we cannot
derive a firm ``no-loose" theorem for stop searches at hadron or $e^+e^-$
colliders. On the other hand, if \st\ is rather light it might well be
long--lived even if the stop--LSP mass difference is large, since for chargino
masses below 100 GeV or so the rule of thumb that the chargino is twice as
heavy as the LSP (for gaugino--like LSP) need not apply. In such a scenario
{\em both} open stop and stoponium production might be observable at hadron
colliders, the former via the $\st \rightarrow c+$LSP loop decay, the latter
in the two--photon channel. Given the intimate connection between stop squarks
and the puzzle of electroweak symmetry breaking, in particular in models where
this breaking occurs radiatively, experimental searches for any signal for
scalar top production at present and future colliders are well worth the
effort.

\subsection*{Acknowledgements}
We thank S. Kim for discussions on preliminary CDF data on two--photon
production, and M. Olsson for discussions on cascade decays of excited
quarkonium states. This work was supported in part by the U.S. Department of
Energy under contract No. DE-AC02-76ER00881, by the Wisconsin Research
Committee with funds granted by the Wisconsin Alumni Research Foundation, as
well as by the Texas National Research Laboratory Commission under grant
RGFY93--221. The work of M.D. was supported by a grant from the Deutsche
Forschungsgemeinschaft under the Heisenberg program.

\renewcommand{\theequation}{A.\arabic{equation}}
\setcounter{equation}{0}

\section*{Appendix}

In this appendix we list the (squared) matrix elements for $S-$wave color
singlet \stst\ pair annihilation into the two--body final states of
eqs.\ref{e7}, as well as the widths for the single stop decays of
eqs.(\ref{e4a},b). We do not include annihilation into two charginos, since
this can only occur if the $\st \rightarrow \tilde{W_1} + b$ decay is
allowed, in which case it swamps all annihilation decays. Recall also that we
assume $\st$ to be the lightest strongly interacting supersymmetric
particle.

We start with the annihilation decays. In the following expressions we have
suppressed color indices, i.e. summation over colors has been performed. The
resulting color factors are included explicitly. As already briefly noted in
sec. 2, the color wave function in the initial state of the matrix element in
eq.(\ref{e5}) is given by $\frac{1}{3} \delta_{ab}$; after contraction with
the color indices of the scattering amplitude, a sum over $a$ and $b$ has to
be taken. Notice that this is {\em not} normalized to unity, which explains
the appearance of the color factor of 3 in eq.(\ref{e5}); however, this
normalization has the practical advantage that annihilation into two
color--singlet particles has color factor one.

\subsection*{ ${\bf  gg, \gamma\gamma}$ final state}

For an $S-$wave initial state, i.e. for $v \rightarrow 0$, these final states
receive contributions only from four--point interactions (fig. 1c); the $t-$
and $u-$channel \st\ exchange diagrams (figs. 1a,b) vanish in this limit (for
physical, i.e. transverse, gauge bosons). The squared $gg$ amplitude can be
written as
\be\label{ap1}
\sum_{\rm color,\ spins}|{\cal M}(\stst \rightarrow gg)|^2_{v=0}
=\frac{32}{9}g_s^4.
\ee
The squared amplitude for the \gamgam\ final state is given by
\be\label{ap2}
\sum_{\rm spins}|{\cal M}(\stst \rightarrow
\gamma\gamma)|^2_{v=0} =16 q^4 e^4,
\ee
where $q=2/3$ is the charge of stop.

\subsection*{$W^+W^-$ final state}
This final state receives contributions from diagrams with $t-$channel exchange
of sbottoms (only $\tilde{b}_L$ contributes if terms $\propto m_b$ are
neglected), $s-$channel exchange of light ($h \equiv H_2$) and heavy ($H
\equiv H_1$) neutral scalar Higgs bosons, and also from the 4 point
interaction of two stops and two $W$ bosons. The $s-$channel exchange of the
$Z$ boson does not contribute to the $S-$wave amplitude. Here we list the
amplitude for specific helicities $\lambda,\bar{\lambda}$ of the $W$ bosons;
$\lambda, \ \bar{\lambda}$ can take the values $0, \ \pm 1$.
\beq \label{ap3}
{\cal M}^{\lambda \bar{\lambda}}(\stst \rightarrow \ww) &=
\gamma_W^{2-\lambda-\bar{\lambda}} \left( \delta_{\lambda 0}
\delta_{\bar{\lambda}0} \beta_W^2 + (-1)^{\lambda} \delta_{\lambda
\bar{\lambda}} \right) \left[
\frac{1}{2}g^2 \cos^2\theta_t
-\sum_i \frac{g_{H_i WW}c_{\tilde{t}_1}^{(i)}}{4\mstsq-m_{H_i}^2} \right]
\nonumber \\
&-2 \beta_W^2 \gamma_W^2 \delta_{\lambda 0} \delta_{\bar{\lambda} 0}
g^2 \cos^2\theta_t \frac{\mst^2} {\mstsq+\msbsq-m_W^2}.
\eeq
The other combinations of helicities do not contribute; this is easily
understand from spin conservation. In eq.(\ref{ap3}) we have introduced
$\gamma_W=\mst/m_W$ and $\beta_W=\sqrt{1-(m_W/\mst)^2}$; $\gamma_W$ is a
kinematical factor which appears in the polarization vector of longitudinal
gauge bosons ($\lambda=0$).

We have included mixing between $SU(2)$ doublet and singlet stops ($\stl$,
$\str$), defined as in eq.(\ref{e2}) of the main text; however, we ignored
sbottom mixing. The $g_{H_i WW }$ are the Higgs \ww\ couplings \cite{28}:
\be\label{ap3p}
g_{H_1 WW}=g m_W \cos (\beta-\alpha), \ \ g_{H_2 WW } = g m_W\sin
 (\beta-\alpha).
\ee
The Higgs \stst\ couplings $c_{\tilde{t}_1}^{i}$ are defined in eqs.(A3)--(A5)
of ref.\cite{36}; we list them here for completeness:
\beq \label{ap4}
c^{(i)}_{\tilde{t}_1} &= \frac{g m_Z} {\cos \! \theta_W} s^{(i)}
( \frac{1}{2} \cos^2 \theta_t - \frac{2}{3} \sin^2 \theta_W \cos \! 2 \theta_t
)
+ \frac {g m_t^2} {m_W} r^{(i)}_u \nonumber \\
& - \frac {g m_t \sin \! 2 \theta_t} {2 m_W} (A_t r_u^{(i)} + \mu
r'^{(i)}_u),
\eeq
where
\ben \label{a5} \beq
s^{(1)} &= - \cos (\alpha + \beta); \ \
s^{(2)} = \sin (\alpha + \beta); \nonumber \\
r^{(1)}_u &= - \frac {\sin \! \alpha}{\sin \! \beta}; \ \
r^{(2)}_u = - \frac {\cos \! \alpha}{\sin \! \beta}; \nonumber \\
r'^{(1)}_u &= - \frac {\cos \! \alpha} {\sin \! \beta}; \ \
r'^{(2)}_u =   \frac {\sin \! \alpha} {\sin \! \beta}.
\label{a5b} \eeq \een
Here \tanb\ is the ratio of vacuum expectation values introduced in sec. 2,
and $\alpha$ is the mixing angle of the neutral scalar Higgs bosons \cite{28}.
Note that $r^{(2)}_u \rightarrow 1$ and $ r'^{(2)}_u \rightarrow \cot \!
\beta$ if the pseudoscalar Higgs boson is much heavier than $m_Z$; the last
term in eq.(\ref{ap4}) is thus proportional to the off--diagonal entry of the
stop mass matrix in this limit, as emphasized in the text.

\subsection*{${\bf ZZ}$ final state}

The contributing Feynmann diagamms are similar to those for the \ww\
final state, except now the $t-$channel exchanges proceed
through $\tilde{t}_{1,2}$, and crossed ($u-$channel) diagrams have to be added
since the $Z$ bosons don't carry a charge. We find:
\beq\label{ap5}
{\cal M}&^{\lambda\bar{\lambda}}(\stst \rightarrow ZZ)_{v=0}
=-\gamma_Z^{2-|\lambda|-|\bar{\lambda}| }
\left(\delta_{\lambda 0}\delta_{\bar{\lambda} 0}\beta_Z^2
+(-)^{\lambda}\delta_{\lambda \bar{\lambda}})\right.\nonumber\\
&\cdot\frac{1}{\cos^2\theta_W}\left[2g^2
\left((\frac{1}{4}-\frac{2}{3}\sin^2\theta_W)\cos^2\theta_t
+\frac{4}{9}\sin^4\theta_W \right)
-\sum_i \frac{g_{H_iWW}c_{\tilde{t}_1}^{(i)}}{ 4\mstsq-m_{H_i}^2 }
\right]\nonumber\\
& +\frac{2g^2\mstsq}{\cos^2\theta_W}\beta_Z^2\gamma_Z^2
\delta_{\lambda 0}\delta_{\bar{\lambda}0}
\left[\frac{\left(\cos^2\theta_t-\frac{4}{3}\sin^2\theta_W\right)^2}
{2\mstsq-m_Z^2}+\frac{\cos^2\theta_t\sin^2\theta_t}{\mstsq+\msttsq-m_Z^2}
\right],\nonumber\\
\
\eeq
where $\gamma_Z=\mst/m_Z$ and $\beta_Z=\sqrt{1-(m_Z/\mst)^2}$.

\subsection*{$\bf Z\gamma$ final state}

Since the photon does not have longitudinal polarization states, the $t-$ and
$u-$channel exchange of $\st$ again disappears in the $v\rightarrow 0$ limit.
Furthermore, only $\lambda_Z= \lambda_{\gamma}=\pm 1$ states are allowed.
After summing over the final state polarization, we get
\be\label{ap6}
\sum_{\rm spin}
| {\cal M}(\stst \rightarrow Z \gamma)|_{v=0}^2=\frac{8 q^2e^2g^2}{\cos^2
 \theta_W}
\left(\frac{1}{2}\cos^2\theta_t-\frac{2}{3}\sin^2\theta_W\right)^2.
\ee
\subsection*{$\bf hh$ final state}
Here all four classes of diagrams depicted in fig.1 contribute:
\beq\label{ap7}
{\cal M}(\stst\rightarrow &hh)|_{v=0}
=\left\{ \frac{2(c^{(2)}_{\tilde{t}_1})^2}{2\mstsq-m^2_{H_2}}+
\frac{2(c^{(2)}_{\tilde{t}_1 \tilde{t}_2})^2}
{\mstsq+\msttsq -m^2_{H_2}}+c^{22}_{11}\right.
\nonumber\\
&+\frac{c_{\tilde{t}_1}^{(1)}}{4\mstsq-m_{H_1}^2}\frac{gm_Z}{2\cos\theta_W}
[2\sin\! 2\alpha \sin(\beta+\alpha)-\cos(\beta+\alpha)\cos\!2\alpha]\nonumber\\
&\left. +  \frac{c^{(2)}_{\tilde{t}_1}}{4\mstsq-m_{H_2}^2}
\frac{3gm_Z}{2\cos\theta_W}
\cos\! 2\alpha\sin(\beta+\alpha )\right\}.
\eeq
\noindent
Here $c_{\tilde{t}_1 \tilde{t}_2}^{(2)}$ and $c_{11}^{22}$ are the
$\st-\stt-h$ and $\st-\st-h-h$ coupling, respectively; they can be
expressed as
\ben\label{ap8}\beq
c_{\tilde{t}_1\tilde{t}_2}^{(2)}
=&g\frac{m_Z\sin(\alpha+\beta)}{\cos\theta_W}\sin\! 2\theta_t
\left(\frac{2}{3}\sin^2\theta_W-\frac{1}{4}\right)
\nonumber\\
&+\frac{g m_t}{2m_W\sin\beta}(A_t\cos\alpha-\mu\sin\alpha)\cos
2\theta_t;\\
%\eeq
%and $c_{11}^{22}$ is $\st$-$\st$-$h$-$h$ coupling
%and
%\be\label{ap9}
c_{11}^{22}=&\frac{g^2}{2}
\left[ \frac{\cos 2\alpha}{\cos^2\theta_W}
(\frac{1}{2}\cos^2\theta_t-\frac{2}{3}\sin^2\theta_W \cos 2\theta_t)-
\frac{m_t^2}{m_W^2}\frac{\cos^2\alpha}{\sin^2\beta}\right].
\eeq\een

\subsection*{$\bf\tilde{Z}_i \tilde{Z}_j$ final state}

\noindent
This process proceeds by $t-$ and $u-$channel exchange of a
top quark and $s-$channel exchange of Higgs bosons. In our convention
neutralino eigenstates are obtained  by diagonalizing
the neutralino mass matrix by a real orthogonal matrix,
thus the mass eigenvalue of a neutralino can
be either positive or negative. Defining $h$ and $\bar{h}$
to be the helicities of the two neutralinos ($h, \ \bar{h} =\pm 1/2$), we
have:
\beq\label{ap10}
{\cal M}(\stst&\rightarrow \tilde{Z}_i\tilde{Z}_j)
=\delta_{h\bar h}\sqrt{ 4m_{ \tilde{t}_1 }^2-
(m_{\tilde{Z}_i}+m_{\tilde{Z}_j})^2 }
\nonumber\\
&\left\{
\frac{2m_t (a_i a_j-b_ib_j)+( m_{\tilde{Z}_i}+ m_{\tilde{Z}_j})(a_ia_j
+b_ib_j)}
{\frac{1}{2}(m_{\tilde{Z}_i}^2 + m_{\tilde{Z}_j}^2)-m_{\tilde{t}_1}^2
-m_t^2}
\right.
\nonumber\\
&+g\left.\left[\frac{ c_{\tilde{t}_1}^{(1)}(\sin\alpha S_{ij}^{''}
-\cos\alpha Q_{ij}^{''}) }{ 4m_{\tilde{t}_1}^2-m_{H_1}^2 }
+\frac{ c_{\tilde{t}_1}^{(2)}(\sin\alpha Q_{ij}^{''}
+\cos\alpha S_{ij}^{''}) }{ 4m_{\tilde{t}_1}^2-m_{H_2}^2 }
\right]\right\}.
\eeq
Here $a_i$ and $b_i$ are scalar and pseudoscalar stop--top--neutralino
couplings; explicit expressions are given in eqs.(3), (8) and (9) of
ref.\cite{36}. $Q_{ij}^{''}$ and $S_{ij}^{''}$ are Higgs--neutralino couplings
defined in ref.\cite{28}; recall that they are real in our notation.

\subsection*{$\bf b\bar{b}$ final state}

This process proceeds via the $t-$channel exchange of charginos ($\tilde{W}_1,
\tilde{W}_2$) as well as $s-$channel exchange of Higgs bosons:
\beq\label{ap11}
{\cal M}(\stst & \rightarrow b\bar{b})
=-\delta_{h\bar{h}}2 \sqrt{3} \sqrt{m_{\tilde{t}_1}^2-m_b^2}
%\nonumber\\
\cdot \nonumber\\
&\left\{
\sum_{i=1}^2 \frac{1}{3}\frac{m_{\tilde{W}_i} ( c_i^2 -d_i^2) +
m_b (c_i^2+d_i^2) }
{m_b^2-m^2_{\tilde{t}_1}-m^2_{\tilde{W}_i}}
\right.
\nonumber\\
&\left.
\ \ \ +\frac{g m_b}{2 m_W \cos\beta}
\left[
-\frac{c_{\tilde{t}_1}^{(1)}\cos\alpha}{4\mstsq-m_{H_1}^2}
+\frac{c_{\tilde{t}_1}^{(2)}\sin\alpha}{4\mstsq-m_{H_2}^2}
\right]
\right\}.
\eeq
\noindent
Here $h$ and $\bar{h}$ are again the final state helicities, and $c_i$ and
$d_i$ are scalar and peudoscalar stop--bottom--chargino couplings defined as
\ben\label{ap12}\beq
{\cal L}_{\tilde{t}_1 \tilde{W}_i b}&=\bar{b}(c_i + d_i \gamma_5 ) \tilde{W}_i
\st + h.c.;\\
c_i&= -\frac{g}{2}V_{i1}\cos\theta_t
+\frac{gm_bU_{i2}}{2\sqrt{2}m_W\cos\beta}\cos\theta_t
+\frac{gm_t V_{i2}}{2\sqrt{2}m_W \sin\beta}\sin\theta_t;\\
d_i&= -\frac{g}{2}V_{i1}\cos\theta_t
-\frac{gm_bU_{i2}}{2\sqrt{2}m_W\cos\beta}\cos\theta_t
+\frac{gm_t V_{i2}}{2\sqrt{2}m_W \sin\beta}\sin\theta_t.
\eeq\een
$U_{ij}$ and $V_{ij}$ are \cite{28} the matrices that diagonalize the chargino
mass matrix. We use the same conventions as in eqs.(A6) to (A.8) of
Ref.\cite{50}; in particular, $m_{\tilde{W}_{1,2}}$ can have either sign.
Notice that $c_i^2-d_i^2$ is proportional to $m_b$; hence the
whole cross section is suppressed by the square of the bottom quark mass. This
decay mode therefore turns out to be negligible for the whole parameter space
of our interests, unless it is ``accidentally" enhanced by an $s-$channel ($h$
or $H$) pole. The factor of $1/3$ for the $\tilde{W}_i$ exchange term is a
color factor necessary for $t-$channel color singlet exchange. We have
included an overall color factor of $\sqrt{3}$, which strictly speaking only
occurs after the (incoherent) summation over the final state color indices.

\subsection*{$\bf t\bar{t}$ final state}

This process proceeds $t-$channel exchange of neutralinos and
gluinos ($\tilde{g}$) and $s-$channel exchange of scalar Higgs bosons:
\beq\label{ap13}
{\cal M}(&\stst\rightarrow t \bar{t})\vert_{v=0}
=- 2 \delta_{h\bar{h}} \sqrt{3}
\sqrt{m_{ \tilde{t}_1 }^2-m_{t}^2 }\nonumber\\
&\cdot\left\{
\frac{1}{3}\sum_{i=1}^4
\frac{m_{\tilde{Z}_i}(a_i^2 - b_{i}^2)+ m_t (a_i^2 +b_i^2)}
{m_t^2 -m_{\tilde{t}_1}^2-m_{\tilde{Z}_i}^2}
+\frac{4}{9}\frac{m_{\tilde{g}}(a_{\tilde{g}}^2-b_{\tilde{g}}^2)
+m_t(a_{\tilde{g}}^2+b_{\tilde{g}}^2)}
{m_t^2-m_{\tilde{t}_1}^2-m_{\tilde{g}}^2}
\right.\nonumber\\
&\ \ \ \ \ \ \ \ \ \ \left. - \frac{g m_t}{2m_W \sin\beta}
\left[\frac{c^{(1)}_{\tilde{t}_1}\sin\alpha}{4m^2_{\tilde{t}_1}-m_{H_1}^2}
+ \frac{c_{\tilde{t}_1}^{(2)} \cos\alpha}{4m_{\tilde{t}_1}-m_{H_2}^2}
\right]\right\}.
\eeq
The couplings $a_i, \ b_i$ have already occured in eq.(\ref{ap10}) above;
$a_{\tilde g}, \ b_{\tilde g}$ are the corresponding gluino--stop--top
couplings:
\beq\label{ap14}
a_{\tilde{g}}^2+ b_{\tilde{g}}^2&=g_3^2;\nonumber\\
a_{\tilde{g}}^2+b_{\tilde{g}}^2&=-g_3^2 \sin 2 \theta_t.
\eeq
Here again the $\tilde{Z}_i$ exchange term receives a color factor of $1/3$,
while the $\tilde{g}$ exchange contribution comes with a factor of $4/9$ for
color octet exchange. Notice that we have again included an overall factor
of $\sqrt{3}$ which properly only appears in the squared amplitude after
summation over the $t \bar{t}$ color states.

\subsection*{Single stop decays: $\st \rightarrow \tilde{W}_i+b,\ \tilde{Z}_j
+t$}
If the stop mass is larger than $m_{\tilde{W}_1}+m_b$ or
$m_{\tilde{Z}_1}+m_t$, single stop decays dominate and the pair annihilation
modes described above all have a very small branching ratio. The decay width
into $b + \tilde{W}_i$ is given by:
\be\label{ap15}
\Gamma(\st\rightarrow b \tilde{W}_i^{+}) =
\frac{|{\cal M}|^2 }{16\pi\mst}
\sqrt{\left(1- \frac{m_b^2+m_{\tilde{W}_i}^2}{\mstsq}
          \right)^2-\frac{4m_b^2m_{\tilde{W}_i}^2}{m_{\tilde{t}_1}^4} }
\ee
where
\be\label{ap16}
\sum_{\rm spin}| {\cal M}|^2=
2c_i^2 \left[ m_{\tilde{t}_1}^2 -(m_b+ m_{\tilde{W}_i})^2 \right]^2
+2 d_i^2\left[ m_{\tilde{t}_1}^2 -(m_b- m_{\tilde{W}_i})^2 \right]^2.
\ee
The couplings $c_i$ and $d_i$ have been defined in Eq.(\ref{ap12}).
Eqs.(\ref{ap15},\ref{ap16}) also describe the decay width for $\st \rightarrow
\tilde{Z}_j + t$, with the following substitutions: $m_b \rightarrow m_t, \
m_{\tilde{W}_i} \rightarrow m_{\tilde{Z}_j}, \ c_i \rightarrow a_j$ and $d_i
\rightarrow b_j$; the $t \st \tilde{Z}_j$ couplings $a_j$ and $b_j$ have
already been introduced in eq.(\ref{ap10}).

\clearpage
\section*{Figure Captions}

\renewcommand{\labelenumi}{Fig.\arabic{enumi}}
\begin{enumerate}

\item %Fig.1
The four classes of Feynman diagrams that contribute to annihilation decays of
stoponium into two--body final states.

\vspace*{5mm}
\item  %Fig.2
Branching ratios for annihilation decays of \sigst\ listed in eq.(\ref{e7}).
The range of \mst\ values shown results from varying $A_t$ between --312 and
--67 GeV. The values of the other parameters are: $\mstl=\mstr=200$ GeV,
$m_t = -\mu = 150$ GeV, $M_2 = 100$ GeV, $m_P = 500$ GeV and \tanb\ = 2.
The branching ratios for the $b \bar{b}$ and $t \bar{t}$ final states (not
shown) are always below 10$^{-3}$.

\vspace*{5mm}
\item  %Fig.3
The binding energy of the lowest stoponium state \sigst\ (dotted) is compared
with the total \sigst\ decay width (solid, dashed), for two different sets
of parameters. We have chosen $m_t = 150$ GeV, $m_P = 500$ GeV, $\tanb = 2, \
\mstl = 400$ GeV, and $\mstr = 300$ GeV. The solid and dashed curves correspond
to scenarios with a gaugino--like and higgsino--like LSP, respectively.

\vspace*{5mm}
\item   %fig.4
Branching ratios for annihilation decays of \sigst\ listed in eq.(\ref{e4}).
The range of \mst\ values shown results from varying $A_t$ between 440 and
1080 GeV. We have increased the $SU(2)$ gaugino mass $M_2$ along with \mst\,
so that the tree--level single stop decays of eq.(\ref{e3}) remain
kinematically forbidden ($M_2 = 1.5 \mst$). The values of the other parameters
are: $m_t = 150$ GeV, \mstl\ = 400 GeV, \mstr\ = $-\mu=$ 300 GeV, $m_P = 500$
GeV, and \tanb\ = 2. The branching ratio for the $b \bar{b}$ mode is again
small.

\vspace*{5mm}
\item  %Fig. 5
Dominant partial widths for \sigst\ annihilation decays. The $gg$ partial
width (dotted) depends only on \mst\ and the QCD scale parameter $\Lambda$,
while the $hh$ (solid, short--dashed) and \ww\ (long--dashed) partial widths
in general depend on all parameters entering the stop mass matrix of
eq.(\ref{e1}). We have kept $A \equiv 2 A_t / (\mstl+\mstr)$ as well as the
ratio \mstl/\mstr\ fixed and varied \mstl. Most curves are for $\mstr=\mstl$,
but the short--dashed curve has been obtained with $\mstr = 0.7 \mstl$. The
values of the other parameters are: $m_t = 150$ GeV, $\tanb = 3, \ \mu = 500$
GeV and $m_P = 1 $ TeV.

\vspace*{5mm}
\item  %Fig. 6
Cross section for $p \bar{p} \rightarrow \sigst \rightarrow \gamgam$ after
cuts at the tevatron. The dashed curve assumes $\Gamma_{\rm tot}(\sigst) =
\Gamma(\sigst \rightarrow gg)$, while the solid line includes all channels
listed in eqs.(\ref{e7}); the difference is noticeable only for $\msig \simeq
m_h$. The dotted curves show our estimates of the minimal signal that is
visible on top of the smooth \gamgam\ background, for three different values
of the integrated luminosity. The signal has been computed for $m_t=150$ GeV,
$\tanb=2, \ M_2=1.5\mst, \ \mstl=1.5 \mstr = 300$ GeV, $m_P = 500$ GeV and
$\mu = -133$ GeV.

\vspace*{5mm}
\item  %Fig. 7
Cross section for \sigst\ production at the LHC. The solid line shows the
total cross section multiplied with 0.01, and the dashed curves the $\gamgam$
signal cross section after cuts, for the two scenarios of figs.2 and 4. The
dotted curve shows the minimal cross section giving a significant signal after
one year of nominal LHC operations, as defined in the text.

\vspace*{5mm}
\item  %Fig. 8
The dependence of the \gamgam\ signal cross section after cuts on the choice
of scale $Q$ in the gluon distribution functions, and on the QCD parameter
$\Lambda$. The parameters are as in fig.5, with $A=1$.

\vspace*{5mm}
\item  %Fig. 9
The ratio $\sum_n |R_n(0)|^2 / |R_1(0)|^2$, where $n$ runs over all true
stoponium bound states, defined by eq.(\ref{e17}). This is a measure of the
possible enhancement of the signal for stoponium production due to the
production of excited states, as discussed in the text.

\vspace*{5mm}
\item  %Fig. 10
The region in the plane spanned by \msig\ and $A_t + \mu \cot \! \beta$ after
one (solid) and five (dashed) years of running the LHC at full luminosity
(${\cal L} = 100$ fb$^{-1}$ per year). The region in the top left corner is
exlcuded by LEP searches for neutral Higgs bosons. The curves have been
obtained for $m_t = 150$ GeV, $\tanb = 3, \ M_2 = 1.67 \mst, \ \mstl=\mstr, \
\mu=750$ GeV and $m_P = 2$ TeV, but depend little on these choices unless
$\msig \simeq m_P$, as discussed in the text.

\end{enumerate}
\end{document}